\documentclass[prb,aps,twocolumn,showpacs]{revtex4}

\usepackage{graphicx,epsfig}
\usepackage{bm}

\begin{document}

\title{Quantum quenches and driven dynamics in a
       single-molecule device}
\smallskip

\author{Yuval Vinkler,$^1$ Avraham Schiller,$^1$
        and Natan Andrei$^2$}

\affiliation{$^1$Racah Institute of Physics,
                 The Hebrew University, Jerusalem 91904,
                 Israel\\
             $^2$Center for Materials Theory, Department
                 of Physics, Rutgers University,
                 Piscataway, NJ 08854-8019 USA}

\begin{abstract}
The nonequilibrium dynamics of molecular devices
is studied in the framework of a generic model for
single-molecule transistors: a resonant level coupled
by displacement to a single vibrational mode. In the
limit of a broad level and in the vicinity of the
resonance, the model can be controllably reduced to
a form quadratic in bosonic operators, which in turn
is exactly solvable. The response of the system to a
broad class of sudden quenches and ac drives is
thus computed in a nonperturbative manner, providing
an asymptotically exact solution in the limit of
weak electron-phonon coupling. From the analytic
solution we are able to
(1) explicitly show that the system thermalizes
    following a local quantum quench,
(2) analyze in detail the time scales involved,
(3) show that the relaxation time in response to
    a quantum quench depends on the observable
    in question, and
(4) reveal how the amplitude of long-time
    oscillations evolves as the frequency of an
    ac drive is tuned across the resonance
    frequency.
Explicit analytical expressions are given for all
physical quantities and all nonequilibrium scenarios
under study.
\end{abstract}

\pacs{73.63.−b, 71.38.−k, 85.65.+h}


\maketitle

\section{Introduction}

The description of strong electronic correlations
far from thermal equilibrium constitutes one of
the major open questions of modern condensed
matter physics. Even under the most favorable
conditions of nonequilibrium steady state, many
of the concepts and techniques that have proven
so successful in equilibrium are simply inadequate.
Recent advancements in a broad range of systems,
from time-resolved
spectroscopies~\cite{Spectroscopy-1,Spectroscopy-2}
to cold atoms~\cite{cold-atoms-1,cold-atoms-2} and
driven nanostructures,~\cite{Elzerman04,Petta05} have
opened new and exciting possibilities for studying
the nonequilibrium dynamics in response to quantum
quenches and forcing fields. Depending on the physical
context one is interested in questions of both
basic and practical nature, such as what are the
underlying time scales governing the dynamics,
how long is coherence maintained, and whether and
how does the system equilibrate at long times.
Some questions, e.g., the issue of equilibration,
often require nonperturbative treatments even if
the system is tuned to weak coupling.

Recent years have witnessed the development of an
array of powerful numerical techniques aimed at
tracking the real-time dynamics of interacting
low-dimensional systems. In the more specific context
of quantum impurity systems these methodologies
include time-dependent variants of the density-matrix
renormalization group,~\cite{Schmitteckert,Feiguin}
the time-dependent numerical renormalization
group,~\cite{TD-NRG-PRL,TD-NRG-PRB} different
continuous-time Monte Carlo
approaches,~\cite{Muehlbacher-rabani-2008,
Weiss-et-al-2008,Werner-et-al,Schiro-Fabrizio-2009}
and sparse polynomial space
representations.~\cite{Alvermann-Fehske-2009}
Despite notable successes, part of these methods are
subject to finite-size effects and discretization
errors, while others are confined to rather short
time scales. Analytical efforts in this realm have
focused mainly on suitable adaptations of perturbative
renormalization-group~\cite{Schoeller,FRG} and
flow-equation~\cite{Kehrein} ideas, which in turn
neglect higher order terms. Exact analytical
solutions, when available, are thus invaluable
both for setting a benchmark and for gaining
unbiased understanding of the underlying physics.
Unfortunately such exact solutions are restricted
at present to very special models whose coupling
constants must be carefully
tuned.~\cite{SH-ac+PB,Kehrein-TL}

In this paper we present an asymptotically exact
solution for the nonequilibrium dynamics of a
single-molecule transistor in response to various
quantum quenches and ac drives. Single-molecule
devices have attracted considerable interest
lately due to the technological promise of
molecular electronics.~\cite{Reviews-of-SMTs}
From a basic-science perspective they offer an
outstanding platform to study the electron-phonon
coupling at the nano-scale. In a typical
molecular bridge, molecular orbitals are coupled
simultaneously to the lead electrons and to the
vibrational modes of the molecule, with the
former degrees of freedom reduced to a single
effective band in the absence of a bias
voltage.~\cite{Glazman-Raikh-88} A minimal
model for an unbiased molecular bridge therefore
consists of a single resonant level coupled by
displacement to a single vibrational mode, as
described by the Hamiltonian of
Eqs.~(\ref{H_full}) and (\ref{H_0}) below.

The spinless Hamiltonian of Eqs.~(\ref{H_full})
and (\ref{H_0}) has been extensively used in
recent years to model single-molecule transistors,
however despite its apparent simplicity it
lacks a complete solution. Conventionally the
model is treated either using perturbation theory
in the electron-phonon coupling when the coupling
is sufficiently weak, or using the Lang-Firsov
transformation~\cite{LF62} and the polaronic
approximation in the limit where tunneling is
sufficiently small. A particularly elegant
nonperturbative solution of the model was
recently devised by D\'{o}ra and
Halbritter,~\cite{Dora_Halbritter_2009} who
noticed that the original electronic Hamiltonian
of Eqs.~(\ref{H_full}) and (\ref{H_0}) can be
mapped onto an exactly solvable bosonic form in
the limit where the electronic level is broad.
Building on prior
results~\cite{Dora_2007,Dora_Gulacsi_2008} for
the related single-impurity Holstein model,
these authors proceeded to compute the
temperature-dependent conductance of the
device under strict resonance conditions.
Since mapping onto the exactly solvable model
is controlled by the smallness of the
electron-phonon coupling $g$ as compared to
the level width $\Gamma$, these results are
expected to be asymptotically exact in the
weak-coupling limit.

In this paper we take the solution one step
further by extending it to the nonequilibrium
dynamics in response to a broad class of
quantum quenches and drives. We explicitly
show that the system thermalizes following a
local quantum quench and analyze in detail
the time scales involved. In particular, we
find that the relaxation time depends on the
observable in question, growing by a factor of
two in going from the phonon occupancy to the
phonon displacement and the electronic occupancy
of the level. This is quite surprising since
unlike the Anderson impurity model, where spin
and charge generally relax on different time
scales,~\cite{TD-NRG-PRL} the phonon occupancy
and displacement pertain to the same degrees of
freedom. A related doubling of frequency occurs
in the long-time response of the phonon occupancy
to an ac drive. These results, as well as others,
are obtained in a fully analytic manner, which is
perhaps the most appealing aspect of our solution.

Before proceeding to actual calculations, two
technical comments are in order. First, some of
the scenarios under consideration in this paper
pertain to a level off resonance with the Fermi
energy, which necessitates the incorporation of
the level energy into the bosonic Hamiltonian.
A nonzero energy level breaks particle-hole
symmetry, an aspect that is missing in the
treatment of D\'{o}ra and Halbritter. Below we
correct their mapping to properly account for
this important point. Second, some of the initial
states to be considered will be nonthermal
states that cannot be treated using the Keldysh
technique. We circumvent this complication by
explicitly constructing the single-particle
eigenmodes of the bosonic Hamiltonian and using
them to propagate the system in time.

The reminder of the paper is organized as follows.
In Sec.~\ref{sec:model} we introduce the model
and its mapping onto a form quadratic in bosonic
operators. The bosonic Hamiltonian is solved in
turn in Sec.~\ref{Sec_exact_diag} by explicitly
constructing its single-particle eigenmodes using
the scattering-state formalism. Technical details
of the solution are relegated to the Appendix. The
next three sections are devoted to three different
quench scenarios: one, Sec.~\ref{sec:interaction-quench},
where the electron-phonon interaction is suddenly
switched on, another, Sec.~\ref{sec:frequency-quench},
where the phonon frequency is abruptly shifted
from its initial value, and lastly,
Sec~\ref{sec:level-quench}, the scenario where
a sudden change is applied to the electronic level.
The case of driven dynamics is addressed in
Sec.~\ref{sec:driven-dynamics}, first in its general
form before turning to an explicit discussion of
ac drives. Finally, we present our conclusions
in Sec.~\ref{sec:conclusions}

\section{The Model and its mapping}
\label{sec:model}

The Hamiltonian we consider is one of the common models
used to describe a single Coulomb-blockade resonance in
molecular devices. It consists of a single spinless
electronic level $d^{\dagger}$ with energy $\epsilon_d$,
which is coupled by displacement to a local vibrational
mode $b^{\dagger}$ with frequency $\omega_0$. The
level is further coupled to a band of spinless
electrons via the hopping matrix element $t$, as
described by the Hamiltonian~\cite{comment-on-hbar}
\begin{eqnarray}
{\cal H} = {\cal H}_0 + \epsilon_d \hat{n}_d
         + \omega_0 b^{\dagger}b
         + g \left(
                    b^{\dagger} + b
             \right)
             \left(
                      \hat{n}_d  - \frac{1}{2}
             \right) ,
\label{H_full}
\end{eqnarray}
with $\hat{n}_d = d^{\dagger} d$ and
\begin{equation}
{\cal H}_0 = \sum_k
                  \epsilon_k c^{\dagger}_k c_k
           + t \sum_k
               \left(
                      c^{\dagger}_k d + d^{\dagger} c_k
               \right) .
\label{H_0}
\end{equation}
Here the combination $\hat{Q} = (b^{\dagger} + b)/\sqrt{2}$
can be thought of as a dimensionless position operator
for the local phonon.

The Hamiltonian defined by Eqs.~(\ref{H_full})
and (\ref{H_0}) has a long history that dates
back to the 1970s, when it was proposed as a
model for the electron-phonon coupling in
mixed-valence compounds.~\cite{SvM75} In the
modern context of nanostructures it is expected
to properly describe the physics of single-molecule
devices away from Coulomb-blockade valleys where a
single unpaired spin resides on the molecule.
Typically the Hamiltonian is treated either
in the weak-coupling limit using perturbation
theory in $g$, or using the Lang-Firsov
transformation~\cite{LF62} and the polaronic
approximation in the limit where $t$ is small.
We shall take a different route and present a
nonperturbative solution to this model which is
asymptotically exact in the limit where
$\Gamma \gg \max\{g, |\epsilon_d|, g^2/\omega_0 \}$.
Our approach is based on the fact that the Hamiltonian
of Eqs.~(\ref{H_full}) and (\ref{H_0}) can be
controllably reduced in this limit to a form
quadratic in bosonic operators which is exactly
solvable. As discussed in the introduction, this
method was first employed in equilibrium by
D\'{o}ra and Halbritter.~\cite{Dora_Halbritter_2009}
Here we exploit this property of the model to
calculate the real-time dynamics following different
quantum quenches and also in response to ac drives.
Accordingly, our presentation begins with the
conversion of the Hamiltonian to a form that is
quadratic in bosonic operators, whose solution is
detailed in turn in Sec.~\ref{Sec_exact_diag}.

Technically, the construction of the bosonic
Hamiltonian proceeds in two steps
(Ref.~\onlinecite{Dora_Halbritter_2009}):
(i) the conversion to a continuum-limit
Hamiltonian and (ii) its subsequent bosonization.
Special care is paid to the parametric form
of the coupling constants that enter the bosonic
Hamiltonian and to the role of the energy level
$\epsilon_d$ which breaks particle-hole symmetry.
The latter energy scale is of particular interest
as it can be controlled experimentally using suitable
gate voltages. In this respect our derivation
exceeds that of D\'{o}ra and Halbritter.

\subsection{Conversion to a continuum-limit Hamiltonian}

Our first goal is to map the
Hamiltonian of Eq.~(\ref{H_full}) onto a
continuum-limit form, where the resonant-level
operator $d^{\dagger}$
is replaced with a suitable field operator. To
this end, we first diagonalize the Hamiltonian term
${\cal H}_0$ using scattering theory to construct
its single-particle eigenmodes. These are conveniently
expressed using the Green function of the level
\begin{equation}
G(z) = \left[
              z - \sum_k \frac{t^2}{z-\epsilon_k}
       \right]^{-1}
\label{G(z)}
\end{equation}
and its associated phases
\begin{equation}
\phi_k = {\rm arg} \left \{
                            G(\epsilon_k - i\eta)
                   \right \} ,
\end{equation}
where the limit $\eta \rightarrow 0^+$ is
implied. Specifically, introducing
the properly normalized fermionic operators
\begin{eqnarray}
\psi^{\dagger}_k &=&
      e^{i\phi_k} c^{\dagger}_k
      + t \left| G(\epsilon_k + i\eta) \right|
\nonumber \\
   && \;\;\;\;\;\;\;\;\;\;\;\;\;\,
      \times
      \left[
             d^{\dagger}
             + \sum_{k'}
                    \frac{t}
                         {\epsilon_k-\epsilon_{k'}+i\eta}
                    c^{\dagger}_{k'}
      \right] ,
\end{eqnarray}
the Hamiltonian term ${\cal H}_0$ can be shown to
take the diagonal form
\begin{equation}
{\cal H}_0 = \sum_k \epsilon_k \psi^{\dagger}_k \psi_k ,
\label{H_0-via-psi_k}
\end{equation}
while $d^{\dagger}$ acquires the mode expansion
\begin{equation}
d^{\dagger} = t \sum_k
                       \left |
                               G(\epsilon_k + i\eta)
                       \right | \psi^{\dagger}_k .
\label{d-via-psi_k}
\end{equation}
Further converting to the continuous energy-shell
operators
\begin{equation}
\tilde{\psi}^{\dagger}_\epsilon =
       \frac{1}{\sqrt{\rho(\epsilon)}}
       \sum_k
              \delta(\epsilon - \epsilon_k)
              \psi^{\dagger}_k
\end{equation}
where $\rho(\epsilon)$ is the conduction-electron
density of states, Eqs.~(\ref{H_0-via-psi_k}) and
(\ref{d-via-psi_k}) become
\begin{equation}
{\cal H}_0 =
      \int_{-D}^{D} \!\epsilon
                    \tilde{\psi}^{\dagger}_\epsilon
                    \tilde{\psi}_\epsilon d\epsilon
\end{equation}
and
\begin{equation}
d^{\dagger} = \int_{-D}^{D}\!
                   \sqrt{\rho_d(\epsilon)}
                   \tilde{\psi}^{\dagger}_\epsilon
                   d\epsilon .
\label{d-via-tilde-psi}
\end{equation}
Here $D$ is the conduction-electron bandwidth and
\begin{equation}
\rho_d(\epsilon) = -\frac{1}{\pi}
        {\rm Im}
                 \left \{
                          G(\epsilon_k + i\eta)
                 \right\}
\label{rho_d}
\end{equation}
is the spectral function associated with the Green
function of Eq.~(\ref{G(z)}).

Our manipulations thus far were exact, independent
of details of the band dispersion $\epsilon_k$. To
make further progress we consider hereafter the
wide-band limit, where the spectral function of
Eq.~(\ref{rho_d}) acquires the Lorentzian form $\pi
\rho_d (\epsilon) = \Gamma/(\epsilon_k^2+\Gamma^2)$
with the hybridization width $\Gamma = \pi \rho(0) t^2$
[$\rho(0)$ is the conduction electrons density
of states at the Fermi energy]. Physically,
$\Gamma$ serves as a new high-energy cutoff for
the integration in Eq.~(\ref{d-via-tilde-psi}).
Since $d^{\dagger}$ is the only electronic degree of
freedom that enters the remaining Hamiltonian terms in
Eq.~(\ref{H_full}), $\Gamma$ acts as a new effective
bandwidth for the electron-phonon coupling. We shall
next exploit this observation to further manipulate
the Hamiltonian of the system.

The Lorentzian cutoff in Eq.~(\ref{d-via-tilde-psi}) is
somewhat inconvenient to deal with. However, its precise
form should not play any role in the desired limit
$\Gamma \gg \max \{ g, |\epsilon_d|, g^2/\omega_0 \}$,
allowing one to adopt a more convenient cutoff
scheme. Indeed, it is useful to replace
$\rho_d (\epsilon)$ in Eq.~(\ref{d-via-tilde-psi})
with a rectangular box profile~\cite{LSA03} that has
the same height at $\epsilon = 0$ and shares the
same characteristic width:
\begin{equation}
\rho_d(\epsilon) \to \frac{1}{\pi\Gamma}
                     \theta( D_d - |\epsilon| )
\label{Box-DOS}
\end{equation}
with
\begin{equation}
D_d = \frac{\pi\Gamma}{2} .
\end{equation}
Substituting $\rho_d(\epsilon)$ with the box profile
of Eq.~(\ref{Box-DOS}), the full Hamiltonian of
Eq.~(\ref{H_full}) becomes
\begin{eqnarray}
{\cal H}\! &=& \!\int_{-D}^{D}\!
                      \epsilon
                      \tilde{\psi}^{\dagger}_{\epsilon}
                      \tilde{\psi}_{\epsilon} d\epsilon
            + \omega_0 b^{\dagger}b
\label{H_w_box-dos} \\
         &+&\! \left [
                       \frac{\epsilon_d}{\pi\Gamma}
                       + \frac{g}{\pi \Gamma}
                       \left(
                              b^{\dagger} + b
                       \right)
                \right ]
                \int_{-D_d}^{D_d}\!d\epsilon\!
                \int_{-D_d}^{D_d}\!d\epsilon'\!
                     :\! \tilde{\psi}^{\dagger}_{\epsilon}
                         \tilde{\psi}_{\epsilon'}\!\!: ,
\nonumber
\end{eqnarray}
where $:\! \tilde{\psi}^{\dagger}_{\epsilon}
\tilde{\psi}_{\epsilon'}\!\!: =
\tilde{\psi}^{\dagger}_{\epsilon}
\tilde{\psi}_{\epsilon'} -
\delta(\epsilon - \epsilon') \theta(-\epsilon)$
stands for normal ordering with respect to the
filled Fermi sea.
Note that all electronic modes with
$|\epsilon| > D_d$ are decoupled from the phonon
in Eq.~(\ref{H_w_box-dos}) and can therefore be
omitted. This amounts to setting $D \to D_d$ in
the integration boundaries of the free
kinetic-energy term.

The conversion to a continuum-limit Hamiltonian is
completed by defining the right-moving field
\begin{equation}
\psi^{\dagger}(x) =
      \frac{1}{\sqrt{2 a D_d}}
      \int_{-D_d}^{D_d}
           e^{-i \epsilon x/v_F}
           \tilde{\psi}^{\dagger}_{\epsilon} d\epsilon ,
\label{Def-right-moving-psi}
\end{equation}
where $v_F$ is the Fermi velocity and
\begin{equation}
a = \frac{\pi v_F}{D_d}
  = \frac{2 v_F}{\Gamma}
\end{equation}
is a new short-distance cutoff corresponding to a
lattice spacing. The new cutoff is connected to the
momentum cutoff $k_{\rm c} = v_F/D_d$ through the
standard relation $k_{\rm c} = \pi/a$. The field
operators so defined obey canonical anticommutation
relations
$\{ \psi(x), \psi^{\dagger}(y) \} = \delta(x - y)$,
subject to the regularization $\delta(0) = 1/a$.
Recalling that the local fermion $d^{\dagger}$
has been mapped in this process onto
$\sqrt{a}\psi^{\dagger}(0)$, this regularization
guarantees that $\{ d, d^{\dagger} \} = 1$ is
preserved. Written in terms of the new field
operators, the Hamiltonian of the system takes the
continuum-limit form
\begin{eqnarray}
{\cal H} \!&=&\! -i v_F
         \int_{-\infty}^{\infty}
              \!\psi^{\dagger}(x) \partial_x \psi(x) dx
         + \omega_0 b^{\dagger}b
\nonumber \\
         &&\!
         + \left [
                   \tilde{\epsilon}_d
                   + \lambda
                   \left(
                          b^{\dagger} + b
                   \right)
           \right ]\!
           :\!\psi^{\dagger}(0) \psi(0)\!:
\label{H_CL}
\end{eqnarray}
with
\begin{eqnarray}
\lambda &=& ga
         = 2\frac{v_F}{\Gamma} g , \\
\tilde{\epsilon}_d &=& \epsilon_d a
         = 2\frac{v_F}{\Gamma} \epsilon_d .
\label{tilde_epsilon}
\end{eqnarray}
Hence, the resonance width $\Gamma$, the
electron-phonon coupling $g$, and the energy level
$\epsilon_d$ have been reduced to two parameters
only, which have the dimension of energy times
length. It should be stressed that the original
conduction-electron bandwidth $D$ has been replaced
in Eq.~(\ref{H_CL}) with $D_d \sim \Gamma$, which
serves as the new high-energy cutoff for the
continuum-limit Hamiltonian.

The Hamiltonian of Eq.~(\ref{H_CL}), first derived
in this context by D\'{o}ra and
Halbritter,~\cite{Dora_Halbritter_2009} is by no
means new. It describes the coupling of a localized
phonon mode to a conduction band, and as such has
been applied in different variants to a broad class
of physical systems. For example, Gadzuk considered
it as a general impurity model~\cite{Gadzuk81} before
applying it to the vibrational line shape of diatomic
adsorbates on metallic clusters.~\cite{Gadzuk92}
Yu and Anderson~\cite{YA84} proposed a closely
related two-band Hamiltonian as a model for the
anomalous properties of A15 materials, while
D\'{o}ra and Gul\'acsi~\cite{Dora_Gulacsi_2008}
used this Hamiltonian to study the inelastic
scattering from local vibrational modes. Although
the model in its general form lacks a full
solution, it can be conveniently handled in the
parameter regime of interest to us using the
methodology of Abelian bosonization.

\subsection{Abelian bosonization}

Our next step is to bosonize the continuum-limit
Hamiltonian defined by Eq.~(\ref{H_CL}). Using
the standard prescriptions of Abelian
bosonization,~\cite{Haldane81} the fermionic
field operator $\psi(x)$ is written as
\begin{equation}
\psi(x) = \frac{1}{\sqrt{2a}}e^{-i\phi(x)} ,
\end{equation}
where the bosonic field $\phi(x)$ has the mode
expansion
\begin{equation}
\phi(x) =
	2\pi i \sum_{q>0}
               \frac{\xi_q}{q}
               \left(
                      a_q e^{iqx} - a^{\dagger}_q e^{-iqx}
               \right)
        - \frac{2\pi x}{L}\!:\!\hat{N}\!\!:\!
        +\,\hat{\theta} .
\end{equation}
Here, $a_q$ and $a^{\dagger}_q$ are canonical bosonic
creation and annihilation operators corresponding to the
Fourier components of the electronic density, $\hat{N}$ is
the total fermionic number operator, $:\!\!\hat{O}\!\!:$
stands for normal ordering with respect to the filled
Fermi sea, and $\hat{\theta}$ is a phase operator
conjugate to $\hat{N}$. The coefficients $\xi_q$ have
the explicit form
\begin{equation}
\xi_q = \sqrt{\frac{q}{2\pi L}}\,e^{-aq/2\pi} ,
\label{xi_q}
\end{equation}
which includes a suitable ultraviolet momentum
cutoff $k_c = \pi/a$.

The rules of bosonization enable one to represent
fermionic operators in terms of bosonic ones with an
important caveat: the bosonized form of the interaction
term is generally not known away from weak coupling.
This uncertainty is removed in the limit of interest
$\Gamma \gg \max \{ g, |\epsilon_d|, g^2/\omega_0 \}$,
when the standard substitution
$:\! \psi^{\dagger}(x) \psi(x) \!: =
(-1/2\pi) \partial_x \phi(x)$ applies. Restricting
attention to this regime, the bosonized Hamiltonian
is thus recast as~\cite{comment-on-k=0}
\begin{eqnarray}
{\cal H} &=&
             \sum_{k>0} \epsilon_k a^{\dagger}_k a_k
          +  \omega_0 b^{\dagger}b
\nonumber \\
         &+& \left[
                    \lambda (b^{\dagger} + b)
                    + \tilde{\epsilon}_d
             \right]
             \sum_{q>0} \xi_q
                        \left(
                               a_q + a^{\dagger}_q
                        \right) .
\label{H_bosonic}
\end{eqnarray}

Another important identity pertains to the
occupancy of the localized electronic level
$\hat{n}_d = d^{\dagger}d$. Since $d^{\dagger}$
has been mapped in the continuum limit onto
$\sqrt{a}\psi^{\dagger}(0)$, then $\hat{n}_d - 1/2$
corresponds to $a\!\!:\!\psi^{\dagger}(0)\psi(0)\!:$,
where we have made use of the fact that the
expectation value of $\psi^{\dagger}(0)\psi(0)$
with respect to the unperturbed Fermi sea is
$1/(2a)$ [see Eq.~(\ref{Def-right-moving-psi})
with $x = 0$]. Accordingly, $\hat{n}_d - 1/2$ has
the bosonized representation~\cite{comment-on-k=0}
\begin{equation}
\hat{n}_d - \frac{1}{2} =
          a \sum_{k > 0}
                 \xi_k (a^{\dagger}_k + a_k) .
\label{n_d-bosonized}
\end{equation}
This identity will play a key role in our
calculations below.

\section{Exact diagonalization}
\label{Sec_exact_diag}

The Hamiltonian of Eq.~(\ref{H_bosonic}) is quadratic
in bosonic operators and as a result is exactly
solvable. In the following section we construct its
single-particle eigenmodes using the scattering-state
formalism. Although of similar technical complexity,
it is advantageous to first address the case where
$\epsilon_d = 0$, and then extend the discussion to
nonzero $\epsilon_d$. This will prove beneficial as
we shall be interested, among other things, in cases
where the level energy is shifted abruptly from
$\epsilon_d = 0$ to nonzero $\epsilon_d$. As we shall
see, such a scenario requires the conversion between
the eigenmodes of the Hamiltonian with and without
$\epsilon_d$. In contrast to the Keldysh technique,
the expansion in terms of the eigenmodes of the
bosonic Hamiltonian will enable us to address cases
of practical interest where the system is initially
prepared in a nonthermal state. For example, if the
phonon initially occupies an excited state. Our
approach is therefore more general than the Keldysh
technique.

\subsection{Scattering states for $\epsilon_d = 0$}

When a free bosonic mode $a^{\dagger}_k$ impinges upon
the local phonon $b^{\dagger}$, it is scattered into a
linear combination of the free bosonic modes and the
localized phonon. This process can be described by the
scattering states $\alpha^{\dagger}_k$, which are
eigenmodes of the bosonic Hamiltonian obeying suitable
boundary conditions of an incoming free particle. The
scattering-state operators can be found by solving the
Lippmann-Schwinger equation, which takes the operator
form
\begin{equation}
[\alpha^{\dagger}_k, {\cal H}] =
	- \epsilon_k \alpha^{\dagger}_k
        + i\eta (a^{\dagger}_k - \alpha^{\dagger}_k) .
\label{Lippmann-Schwinger}
\end{equation}
The role of $\eta \rightarrow 0^+$ in
Eq.~(\ref{Lippmann-Schwinger}) is to guarantee
appropriate boundary conditions. It does not
enter any physical quantities.

A detailed solution of Eq.~(\ref{Lippmann-Schwinger})
is presented in Appendix~\ref{app_scat_state}, using
the methodology developed in Ref.~\onlinecite{SH98}.
Here we quote only the end result. The scattering-state
operators are given by
\begin{eqnarray}
\alpha^{\dagger}_k \!&=&\! a^{\dagger}_k
       + \lambda g(\epsilon_k+i\eta) \xi_k
         \bigg[
                (\epsilon_k - \omega_0) b
                + (\epsilon_k + \omega_0) b^{\dagger}
\nonumber \\
    \!&+&\! 2\omega_0 \lambda
            \sum_{q>0} \xi_q\!
                 \left(
                        \frac{a^{\dagger}_q}
                             {\epsilon_k\!+\!i\eta\!
                              -\!\epsilon_q}
                        + \frac{a_q}
                               {\epsilon_k\!+\!i\eta\!
                                +\!\epsilon_q}
                  \right)\!
         \bigg] ,
\label{eq_sc_state}
\end{eqnarray}
where
\begin{equation}
g(z) = \frac{1}
            {z^2 - \omega_0^2 - 2\omega_0\Sigma(z)}
\label{eq_g_defined}
\end{equation}
is related to the phononic Green function of
Eq.~(\ref{Phonon-GF}) and
\begin{equation}
\Sigma(z) = \lambda^2\sum_{k>0} \xi_k^2
                 \left(
                        \frac{1}{z - \epsilon_k}
                        - \frac{1}{z + \epsilon_k}
                 \right)
\label{Sigma_def}
\end{equation}
is the corresponding self-energy. Both $\Sigma(z)$ and
$g(z)$ are analytic in the upper and lower halves of
the complex plane, have a branch cut along the real axis,
and are even functions of $z$ [i.e., $g(z) = g(-z)$ and
likewise for $\Sigma(z)$]. In addition $g(z^*) = g^*(z)$
and $\Sigma(z^*) = \Sigma^*(z)$. These analytical
properties are useful in establishing some of the
operator identities that will be employed in this paper.
In particular, it can be explicitly shown that, in the
limit where $L \rightarrow \infty$, $\eta \rightarrow 0^+$
and yet $L \eta \rightarrow \infty$, the Hamiltonian
takes the diagonal form
\begin{equation}
{\cal H} =
      \sum_{k>0} \epsilon_k \alpha^{\dagger}_k \alpha_k ,
\label{eq_ham_in_diag_basis}
\end{equation}
while the scattering-state operators maintain
canonical commutation relations:
\begin{equation}
[\alpha_k, \alpha^{\dagger}_q ] = \delta_{k,q} ,
\end{equation}
\begin{equation}
[ \alpha_k,\alpha_q ] = 0 .
\end{equation}
In fact, the latter commutation relations apply
to any finite $\eta$, though only the limit
$\eta \to 0^+$ is of interest to us.

One particularly useful identity is the expansion
of the local phonon mode $b^{\dagger}$ in terms of
the scattering-state operators:
\begin{eqnarray}
b^{\dagger} &=& \lambda
                \sum_{k>0} \xi_k
                     \bigg[
                            g(\epsilon_k - i\eta)
                             (\epsilon_k + \omega_0)
                              \alpha^{\dagger}_k
\nonumber \\
               && \;\;\;\;\;\;\;\;\;\;\;\;\;\; -
                            g(\epsilon_k + i\eta)
                             (\epsilon_k - \omega_0)
                             \alpha_k
                     \bigg] .
\label{b_via_sc_state}
\end{eqnarray}
Combined with the diagonal form of the Hamiltonian of
Eq.~(\ref{eq_ham_in_diag_basis}), one can immediately
write down the time evolution of $b^{\dagger}(t)$ in
the Heisenberg representation, which reads
\begin{eqnarray}
b^{\dagger}(t)\!&=&\!\lambda
                   \sum_{k>0} \xi_k
                        \bigg[
                               g(\epsilon_k - i\eta)
                                (\epsilon_k+\omega_0)
                                e^{i\epsilon_k t}
                                \alpha^{\dagger}_k
\nonumber \\
               && \;\;\;\;\;\;\;\;\;\;\;\;\; -
                               g(\epsilon_k + i\eta)
                                (\epsilon_k - \omega_0)
                                e^{-i\epsilon_k t}
                                \alpha_k
                        \bigg] .
\label{b_time_dep}
\end{eqnarray}
A similar identity applies to the occupancy of the
localized electronic level, whose bosonized form has
been detailed in Eq.~(\ref{n_d-bosonized}). Expanding
the right-hand side of Eq.~(\ref{n_d-bosonized}) as
\begin{equation}
a \sum_{k>0}
        \xi_k (\epsilon_k^2-\omega_0^2)
        \left[
               g(\epsilon_k - i\eta) \alpha^{\dagger}_k
               + g(\epsilon_k + i\eta) \alpha_k
        \right] ,
\label{n_d-via-alpha}
\end{equation}
one has that
\begin{eqnarray}
\hat{n}_d(t) &=&
        \frac{1}{2} +
        a \sum_{k > 0}
               \xi_k (\epsilon_k^2 - \omega_0^2)
               \bigg [
                       g(\epsilon_k - i\eta)
                       e^{i\epsilon_k t}
                       \alpha^{\dagger}_k
\nonumber \\
        && \;\;\;\;\;\;\;\;\;\;\;\;\;\;\;\;
                     + g(\epsilon_k + i\eta)
                       e^{-i\epsilon_k t}
                       \alpha_k
               \bigg] .
\label{n_d_time_dep}
\end{eqnarray}

The operator identities listed in Eqs.~(\ref{b_time_dep})
and (\ref{n_d_time_dep}) are central to our study as they
allow one to track the nonequilibrium dynamics of the
phonon mode and the level occupancy, respectively.
Accordingly, they will be heavily used throughout the
paper.

\subsection{Extension to nonzero $\epsilon_d$}
\label{subsec:nonzero-ed}

As stated above, the inclusion of a nonzero
$\epsilon_d$ is quite straightforward and does not add
to the complexity of computing the scattering-state
operators. Since $\epsilon_d$ adds a term linear
in bosonic operators to the Hamiltonian
[see Eq.~(\ref{H_bosonic})], the resulting
scattering-state operators differ by a simple
$k$-dependent displacement from their $\epsilon_d = 0$
counterparts (see Appendix~\ref{app_scat_state}
for a detailed derivation). Reserving the notation
$\alpha^{\dagger}_k$ for the scattering-state
operators when $\epsilon_d = 0$ and denoting
the new operators by $\beta^{\dagger}_k$,
the latter are given by
\begin{equation}
\beta^{\dagger}_k = \alpha^{\dagger}_k
       + \tilde{\epsilon}_d \xi_k
         \frac{\epsilon_k^2 - \omega_0^2}
              {\epsilon_k + i\eta}
              g(\epsilon_k + i\eta) ,
\label{beta-via-alpha}
\end{equation}
where $\tilde{\epsilon}_d$ and $\alpha^{\dagger}_k$
are specified in Eqs.~(\ref{tilde_epsilon}) and
(\ref{eq_sc_state}), respectively. The shift
in scattering-state operators carries over to
physical observables as well. For example, the
the local phonon mode is expanded as
\begin{equation}
b^{\dagger} = \tilde{b}^{\dagger}
     + \frac{\tilde{\epsilon}_d}{\lambda}
       \omega_0 g(-i\eta)\Sigma(-i\eta) ,
\label{b-for-nonzero_ed-I}
\end{equation}
where $\tilde{b}^{\dagger}$ is given by the same
formal expression of Eq.~(\ref{b_via_sc_state}),
but with $\alpha^{\dagger}_k$ and $\alpha_k$
replaced with $\beta^{\dagger}_k$ and $\beta_k$,
respectively:
\begin{eqnarray}
\tilde{b}^{\dagger} &=& \lambda
                \sum_{k>0} \xi_k
                     \bigg[
                            g(\epsilon_k - i\eta)
                             (\epsilon_k + \omega_0)
                              \beta^{\dagger}_k
\nonumber \\
               && \;\;\;\;\;\;\;\;\;\;\;\;\;\; -
                            g(\epsilon_k + i\eta)
                             (\epsilon_k-\omega_0)
                             \beta_k
                     \bigg] .
\label{b_zero_via_sc_state}
\end{eqnarray}
As discussed below [see Eq.~(\ref{Expr-for-Sigma})
with $\xi \to 0$], the self-energy $\Sigma(-i\eta)$
takes the particularly compact form $-g^2/(\pi \Gamma)$,
hence Eq.~(\ref{b-for-nonzero_ed-I}) can be rewritten
as
\begin{equation}
b^{\dagger} = \tilde{b}^{\dagger}
     + \frac{\epsilon_d}{\pi \Gamma}
       \frac{g}{\omega_0}
       \frac{1}{1 - 2 g^2/(\pi \omega_0 \Gamma)} ,
\label{b-for-nonzero_ed-II}
\end{equation}
where we have expressed the constant shift in terms
of the original model parameters that appear in
Eq.~(\ref{H_full}).

An analogous expansion applies to the occupancy of the
localized level, $\hat{n}_d$, which is written as
\begin{equation}
\hat{n}_d = \tilde{n}_d
          - \frac{a\tilde{\epsilon}_d}{\lambda^2}
            \omega_0^2 g(-i\eta)\Sigma(-i\eta) .
\end{equation}
Here
\begin{equation}
\tilde{n}_d = \frac{1}{2} +
a \sum_{k>0}
        \xi_k (\epsilon_k^2-\omega_0^2)
        \left[
               g(\epsilon_k - i\eta) \beta^{\dagger}_k
               + g(\epsilon_k + i\eta) \beta_k
        \right]
\label{n_d-tilde-via-beta}
\end{equation}
is the same formal expression of Eq.~(\ref{n_d_time_dep})
with the time $t$ set to zero and with $\alpha^{\dagger}_k$
and $\alpha_k$ replaced by $\beta^{\dagger}_k$ and
$\beta_k$, respectively. As with $b^{\dagger}$, one
can exploit the explicit expression for the self-energy
$\Sigma(-i\eta)$ to recast $\hat{n}_d$ in the form
\begin{equation}
\hat{n}_d = \tilde{n}_d
          - \frac{{\epsilon}_d}{\pi \Gamma}
            \frac{1}{1 - 2 g^2/(\pi \omega_0 \Gamma)} .
\label{n_d-nonzero-ed}
\end{equation}

Note that the displacement of the scattering-state
operators and the associated shifts in the
expansions of $b^{\dagger}$ and $\hat{n}_d$ have
a simple physical origin: they reflect the breaking
of particle-hole symmetry in the original Hamiltonian
of Eq.~(\ref{H_full}) inflicted by a nonzero
$\epsilon_d$. This important aspect of $\epsilon_d$
is absent in the mapping of D\'{o}ra and
Halbritter,~\cite{Dora_Halbritter_2009} who accounted
for this energy scale by a simple Lorentzian reduction
of the coupling constant $\lambda$. Some of the
results presented in this work would be missed
out unless the breaking of particle-hole is
properly treated.

Armed with the single-particle eigenmodes of
the full Hamiltonian and with the expansions
of physical operators in terms of these modes,
we are now in position to compute the real-time
dynamics of the system in response to various
quantum quenches and ac drives. Specifically, we
shall consider three quench scenarios: one where
the electron-phonon interaction is abruptly switched
on, another where the phonon frequency is suddenly
shifted from $\omega_0$ to $\omega_0 + \delta \omega$,
and finally a sudden change in the level energy from
$\epsilon_d = 0$ to nonzero $\epsilon_d$. In addition,
we shall consider two ac drives --- one applied to the
local phonon and another applied to the electronic
level. Of particular interest are the characteristic
time-scales that govern the nonequilibrium dynamics
and their dependences on the physical parameters of
the system. These aspects will be analyzed in detail
below.

\section{Switching on the interaction}
\label{sec:interaction-quench}

We begin our discussion with the nonequilibrium
dynamics following an abrupt switching on of the
electron-phonon interaction $g$. We consider the
following scenario. At time $t < 0$ the system is
free of interactions (i.e., $g = 0$), and occupies
a state that is a direct product of the electronic
ground state (the filled Fermi sea)
and an arbitrary phononic state. Typically
one is interested in cases where the phonon has
either a well-defined occupation number $n$ or
resides in a coherent state, though our discussion is
not restricted to these particular choices. At time
$t = 0$ the electron-phonon interaction is abruptly
switched on and the system evolves under the full
Hamiltonian ${\cal H}$. This acts to entangle the
phononic and electronic degrees of freedom, which
are no longer independent. We concentrate our
discussion on zero temperature, yet the derivation
presented below can readily be extended to any
finite temperature $T$ of the Fermi sea.

Formally, the time evolution of the expectation
value of an observable $\hat{O}$ is given by
the standard expression
\begin{equation}
O(t) = \langle \psi_0|
               U^{\dagger}(t,0)\hat{O} U(t,0)
       |\psi_0\rangle ,
\end{equation}
where $|\psi_0\rangle$ is the initial state of the
system and $U(t,0)$ is the time-evolution operator.
One is therefore interested in the expectation
value of $\hat{O}$ in its Heisenberg representation
$\hat{O}(t) = U^{\dagger}(t,0) \hat{O} U(t,0)$
with respect to the initial state $|\psi_0\rangle$.
In the scenario under consideration the initial state
has a simple representation in terms of the eigenmodes
of the initial Hamiltonian with $g = 0$, whereas
the time evolution has a natural representation
in terms of the eigenmodes of the full (i.e.,
final) Hamiltonian ${\cal H}$. Therefore, the
general strategy for calculating $O(t)$ proceeds
as follows. First $\hat{O}(t)$ is represented in
terms of the eigenmodes of the full Hamiltonian
where its time evolution can easily be implemented,
next it is recast in terms of the eigenmodes of the
initial Hamiltonian, and finally the expectation
value with respect to $|\psi_0\rangle$ is evaluated.
Below we implement this procedure to track the
time evolution of the phononic occupancy
$n_b(t) = \langle b^{\dagger} (t) b (t) \rangle$
and displacement $Q(t) = \frac{1}{\sqrt{2}}
\langle b^{\dagger}(t) + b(t) \rangle $. Throughout
the section we set $\epsilon_d$ equal to zero,
corresponding to a level at resonance with the
Fermi energy.

\subsection{Time evolution of phononic operators}

Our first goal is to express $b^{\dagger}(t)$ in
terms of $a_k$, $a^{\dagger}_k$, $b$ and $b^{\dagger}$,
which are the eigenmodes of the initial Hamiltonian
with $g = 0$. The expansion of $b^{\dagger}(t)$ in
terms of the eigenmodes of the final Hamiltonian is
detailed in Eq.~(\ref{b_time_dep}). Substituting the
explicit expression for the scattering-state operators,
Eq.~(\ref{eq_sc_state}), into Eq.~(\ref{b_time_dep})
one obtains
\begin{widetext}
\begin{equation}
b^{\dagger}(t) = \lambda
          \sum_{k > 0} \xi_k
               \left[
                      F(\epsilon_k - i\eta,t)
                      a^{\dagger}_k
                      + F(-\epsilon_k - i\eta,t)
                      a_k
               \right]
           + I_1(t) b
           + I_2(t)b^{\dagger} ,
\end{equation}
where we have introduced three auxiliary functions
\begin{eqnarray}
&& I_1(t) = \lambda^2
            \sum_{k > 0} \xi_k^2
                 \left|
                        g(\epsilon_k + i\eta)
                 \right|^2
                 (\epsilon_k^2 - \omega_0^2)
                 \left(
                        e^{i\epsilon_k t}
                        - e^{-i\epsilon_k t}
                 \right) ,
\\
&& I_{2}(t) = \lambda^2
              \sum_{k > 0} \xi_k^2
                   \left|
                          g(\epsilon_k + i\eta)
                   \right|^2
                   \left[
                          (\epsilon_k + \omega_0)^2
                           e^{i\epsilon_k t}
                          - (\epsilon_k - \omega_0)^2
                             e^{-i\epsilon_k t}
                   \right] ,
\\
&& F(z,t) = g(z)(z + \omega_0)e^{izt}
            + 2 \omega_0 \lambda^2
              \sum_{k > 0} \xi_k^2
                   \left|
                          g(\epsilon_k + i\eta)
                   \right|^2
                   \left(
                          \frac{\epsilon_k + \omega_0}
                               {\epsilon_k - z}
                           e^{i\epsilon_k t}
                          - \frac{\epsilon_k - \omega_0}
                                 {\epsilon_k + z}
                             e^{-i\epsilon_k t}
                   \right) .
\label{eq_f_defined}
\end{eqnarray}
\end{widetext}

In general, one must resort to numerical integration
to accurately evaluate the three functions defined
above at arbitrary time $t$. Results of such numerical
calculations will be presented below for the relevant
observables of interest. It is instructive, however,
to first gain analytical insight by analyzing the
long-time behaviors of the auxiliary functions.
In the limit $L \to \infty$ one can replace the
sums over $k$ with integrals over energy, resulting
in an exponential decay at long times of all
items but the first term on the right-hand side
of Eq.~(\ref{eq_f_defined}). To see this important
point consider $I_1(t)$, for example. Converting the
sum over $k$ into integration over energy, one is left
with the integral
\begin{eqnarray}
&& I_1(t) = (\rho_0\lambda)^2
            \int_{0}^{\infty}\!\!d\epsilon
                 \left|
                        g(\epsilon + i\eta)\right|^2
                        (\epsilon^2 - \omega_0^2)
\nonumber \\
&& \;\;\;\;\;\;\;\;\;\;\;\;\;\;\;
   \;\;\;\;\;\;\;\;\;\;\;\;\;
          \times \epsilon
                 \left(
                        e^{i \epsilon t}
                        - e^{-i\epsilon t}
                 \right)
                 e^{-\epsilon/D_d} ,
\label{i1_integral-1}
\end{eqnarray}
where $\rho_0 = 1/(2\pi v_F)$ is the density of states
per unit length. Focusing on $t \gg 1/D_d$, one may
(i) omit the exponential cutoff $e^{-\epsilon/D_d}$
in Eq.~(\ref{i1_integral-1}) and (ii) interchange
$\epsilon \to -\epsilon$ in the second term in the
parenthesis to obtain
\begin{equation}
I_1(t) \simeq (\rho_0\lambda)^2
               \int_{-\infty}^{\infty}\!d\epsilon
                     \left|
                            g(\epsilon + i\eta)
                     \right|^2
                     \epsilon\,
                     (\epsilon^2 - \omega_0^2)
                     e^{i\epsilon t} .
\label{i1_integral-2}
\end{equation}
The function $g(\epsilon + i\eta)$ is analytic in the
upper half of the complex plane, whereas the analytic
continuation of $g^*(\epsilon + i\eta)$ to the upper
half plane has a set of isolated
poles~\cite{comment-on-poles} of the form
$p_j = \omega_j + i/\tau_j$ with $\tau_j > 0$. Using
these poles one can formally perform the integral in
Eq.~(\ref{i1_integral-2}) to arrive at
\begin{equation}
I_1(t) \simeq 2 \pi i (\rho_0\lambda)^2
       \sum_j R_j (p_j^2 - \omega_0^2) p_j
              e^{i \omega_j t-t/\tau_j} ,
\end{equation}
where $R_j$ is the residue of $|g(\epsilon + i\eta)|^2$
at $p_j$. Thus, for $t \gg 1/D_d$, the
function $I_1(t)$ is well approximated by a discrete
sum of exponential terms that contain both an
oscillatory component and a part that decays in time.
Asymptotically only those terms with the largest
decay time $\tau_j$ dominate, hence $I_1(t)$ closely
follows a simple exponential decay with superimposed
oscillations. A similar procedure can be applied to
$I_2(t)$ and to the term involving the sum over $k$
in the expression for $F(z,t)$, both of which are
found to be dominated by the same set of poles $p_j$
provided $z$ lies in the lower half plane (as is
the case throughout our calculations).

Next we address the poles $p_j$, which are given
by the solutions to the equation
\begin{equation}
z^2 - \omega_0^2 - 2\omega_0\Sigma^{(+)}(z) = 0 ,
\label{Poles-def}
\end{equation}
where $\Sigma^{(+)}(z)$ is the analytic continuation
of $\Sigma^*(\epsilon + i\eta)$ to the upper half
plane. For $L \to \infty$, the self-energy of
Eq.~(\ref{Sigma_def}) has the explicit analytic
expression
\begin{equation}
\Sigma(z) = (\rho_0 \lambda)^2 D_d
            \left [
                    \xi e^{\xi}
                    E_1 (\xi)
                    - \xi e^{-\xi}
                      E_1 (-\xi)
                    - 2
            \right ] ,
\label{Expr-for-Sigma}
\end{equation}
where $\xi$ equals $z/D_d$ and $E_1(z)$ is the
Exponential Integral function.~\cite{AS-E1}
Expanding $E_1(z)$ as a logarithm plus a power
series in $z$ one obtains
$\Sigma^*(\epsilon + i\eta) =
(\rho_0 \lambda)^2 D_d
\left [ i \pi \tilde{\epsilon} - 2 +
{\cal O} \left ( \tilde{\epsilon}^2
\ln \tilde{\epsilon} \right) \right]$
with $\tilde{\epsilon} = \epsilon/D_d$,
resulting in
\begin{equation}
\Sigma^{(+)}(z) = (\rho_0 \lambda)^2 D_d
       \left [
               i \pi \xi - 2 +
               {\cal O} \left (
                                \xi^2 \ln \xi
                        \right)
       \right] .
\end{equation}
In general, Eq.~(\ref{Poles-def}) lacks an analytical
solution. However, in the desired limit where
$\rho_0 \lambda = g/(\pi \Gamma) \ll 1$ and
$\omega_0 \ll D_d$ one can truncate the expansion
of $\Sigma^{(+)}(z)$ at linear order in $\xi$,
to be left with a simple quadratic equation in
Eq.~(\ref{Poles-def}). In this limit only two poles
exist, which differ in the sign preceding the
frequency: $p_{\pm} = \pm\omega + i/\tau$ with
\begin{equation}
\omega = \omega_0
         \sqrt{
                1 - \frac{2}{\pi}
                    \frac{g^2}{\omega_0\Gamma}
                - \frac{1}{\pi^2}
                  \left (
                         \frac{g}{\Gamma}
                  \right )^4
              } .
\label{omega}
\end{equation}
The decay time $\tau$ is given in this approximation
by $\tau = \pi \Gamma^2/(\omega_0 g^2)$, where we have
converted back to the original model parameters of
Eq.~(\ref{H_full}) in writing both the frequency
$\omega$ and the single relaxation time $\tau$. Note
that these expressions for $\omega$ and $\tau$
coincide with second-order perturbation theory in $g$
when applied directly to the electronic Hamiltonian
of Eq.~(\ref{H_full}),~\cite{EGS-preprint} thus
validating the cutoff scheme used in bosonization.
The expression for $\tau$ can be further improved by
going to the next order in $\omega_0/D_d$, i.e.,
by including one more order in $\xi$ in the
expansion of $\Sigma^{(+)}(z)$. This in turn yields
\begin{equation}
\tau = \frac{\pi}{\omega_0}
       \left (
               \frac{\Gamma}{g}
       \right )^2
       \left [
               1 + \frac{2}{\pi}
                   \frac{\omega_0}{\Gamma}
       \right ] ,
\label{tau}
\end{equation}
where we have restricted ourselves to linear order
in $\omega_0/D_d$ in writing the expression in the
square brackets.

As we shall confirm by explicit numerical
calculations, the nonequilibrium dynamics of all
observables of interest is governed exclusively
by $\omega$ and $\tau$ at time scales exceeding
$1/D_d$. Similar results for $\omega$ and
$1/\tau$ were reported by D\'{o}ra and
Halbritter~\cite{Dora_Halbritter_2009} (corresponding
in their notation to the real and imaginary parts
of $\omega_{p \pm}$), yet their expression for
$\omega$ contained the bare conduction-electron
bandwidth $D$ rather than the renormalized one
$D_d \sim \Gamma \ll D$.
Indeed, Eqs.~(\ref{omega}) and (\ref{tau}) are
free of the bandwidth $D$, indicating that one
can safely implement the limit $D \to \infty$
for the Hamiltonian of Eq.~(\ref{H_full}),
provided $\Gamma$, $g$, and $\omega_0$ are all
held fixed. Physically this reflects the fact
that the local phonon couples to the conduction
band by way of the resonant level only, hence its
level width $\Gamma$ serves as a new effective
high-energy cutoff. By contrast, there is
no meaningful $D_d \to \infty$ limit for the
continuum-limit Hamiltonian of Eq.~(\ref{H_CL})
that keeps both $\omega$ and $\tau$ finite.

Equation~(\ref{omega}) features two special values
of the electron-phonon coupling $g$. One, $g_0$,
above which the frequency $\omega$ becomes imaginary
(i.e., $p_{\pm}$ become purely imaginary) and
another, slightly larger coupling $g_c$, above which
the pole $p_{-}$ is shifted to the lower half plane.
The former coupling strength represents the point
where the local phonon is completely softened, whereas
the latter value represents the point above which
the energy of the lowest bosonic eigenmode of the
Hamiltonian of Eq.~(\ref{H_bosonic}) becomes negative.
Both values of $g$ lie well beyond the applicability
of our theory, as the mapping onto the
continuum-limit Hamiltonian assumed
$\Gamma \gg \max\{g, g^2/\omega_0\}$. Interestingly,
it has been argued by D\'{o}ra~\cite{Dora_2007} that
the bosonized Hamiltonian of Eq.~(\ref{H_bosonic})
offers a faithful representation of the continuum-limit
Hamiltonian of Eq.~(\ref{H_CL}) all the way up to
strong coupling, where the nonlinear conversion
between the fermionic and the bosonic coupling
constants is not explicitly known. In
particular, the point where $p_{-}$ is shifted
to the lower half plane was identified by D\'{o}ra
with $\lambda \to \infty$. It remains to be seen
whether such strong electron-phonon couplings can
indeed be described by a bosonized Hamiltonian
with just a simple linear displacement coupling,
or whether additional nonlinear terms must be
included.

\subsection{Phononic occupancy and displacement}
\label{sec:Q1-nb}

With the explicit expansions of $b^{\dagger}(t)$
and $b(t)$ at hand we can proceed to compute the
time evolution of physical observables, starting
with the phonon occupancy $n_b(t)$ and
displacement $Q(t)$. Since $b^{\dagger}(t)$ is
linear in the eigenmodes of the initial
Hamiltonian, the phonon number operator
$\hat{n}_b(t) = b^{\dagger}(t)b(t)$ is quadratic
in these operators. When averaged with respect
to the initial state, only the combinations
$a_k a^{\dagger}_k$, $b^{\dagger}b$,
$b b^{\dagger}$, $bb$, and
$b^{\dagger}b^{\dagger}$ contribute to the
expectation value of $\hat{n}_b$ at time $t$,
resulting in
\begin{eqnarray}
n_b(t)\!&=&\!\lambda^2
           \sum_{k>0} \xi_k^2
                \left |
                        F(-\epsilon_k - i\eta,t)
                \right |^2
                + \left| I_1(t) \right|^2
                  [n_b(0) + 1]
\nonumber \\
      &&
              \!+ \left | I_2(t) \right|^2 n_b(0)
                + 2{\rm Re} \left \{
                                     I_1(t)I_2^*(t)
                                     \langle
                                             bb
                                     \rangle_{t=0}
                            \right \} .
\label{n_b(t)-sc-1}
\end{eqnarray}

Equation~(\ref{n_b(t)-sc-1}) is the central result
of this subsection. It provides an asymptotically exact
expression for the time evolution of $n_b(t)$ in the
weak-coupling regime. Several comments should be made
about this result. First, the occupancy $n_b(t)$
depends on the initial state of the phonon via two
parameters only: $n_b(0)$ and $\langle bb \rangle_{t=0}$.
Any two initial states that share the same values of
$n_b(0)$ and $\langle bb \rangle_{t=0}$ will produce
identical curves for $n_b(t)$. Second, since $I_1(t)$
and $I_2(t)$ decay to zero with time, the occupancy
at long times is independent of the initial state of
the phonon. Third, the term involving the summation
over $k$ in Eq.~(\ref{eq_f_defined}) decays to zero
as well, resulting in a compact expression for the
phononic occupancy at long times:
\begin{equation}
n_b(t \to \infty) = \lambda^2
          \sum_{k>0} \xi_k^2
               \left |
                       g(\epsilon_k + i\eta)
               \right|^2
               (\epsilon_k - \omega_0)^2 .
\label{eq_occ_infty}
\end{equation}
Finally, one can show that Eq.~(\ref{eq_occ_infty})
is just the zero-temperature equilibrium phonon
occupancy with respect to the full
Hamiltonian,~\cite{comment-on-equilibrium-occ}
implying thermalization at long times. This result
on its own is not surprising, since it has been
rigorously shown by Ambegaokar~\cite{Ambegaokar_2007}
that Hamiltonians involving a local bosonic mode coupled
linearly to a macroscopic bosonic bath do indeed
equilibrate at long times in response to a local
quantum quench. Below we analyze in detail the decay
to the new thermal equilibrium.

\begin{figure}[tb]
\centerline{
\includegraphics[width=80mm]{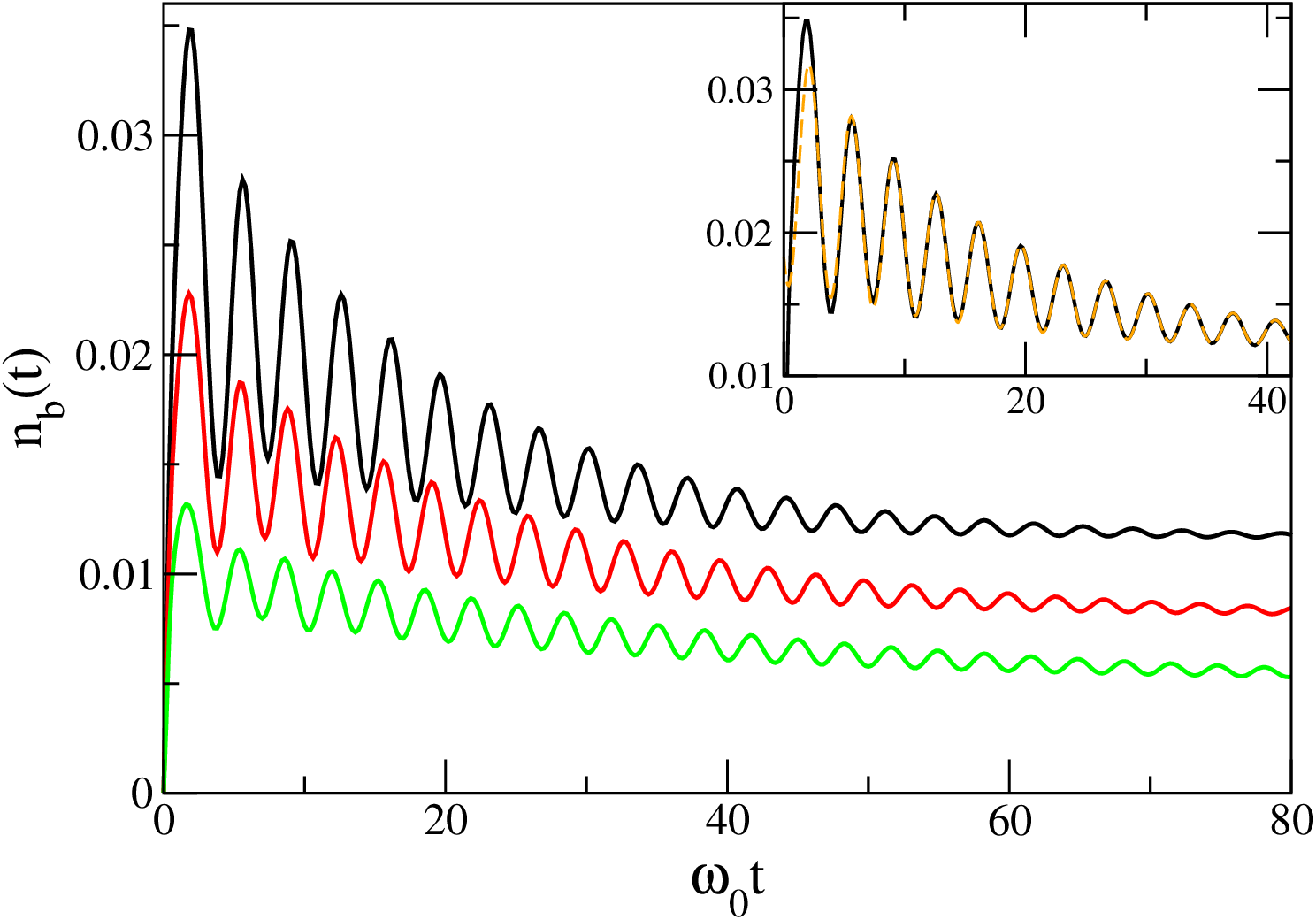}
}\vspace{0pt}
\caption{(Color online)
         Time evolution of the phonon occupancy
         $n_b(t)$ following an abrupt switching on
         of the electron-phonon coupling $g$ at time
         $t = 0$, with the phonon initially occupying
         the empty state $| 0\rangle$. Here
         $\omega_0/D_d = 0.2$, while $g/\Gamma$ equals
         $0.229$ (green), $0.28$ (red), and $0.324$
         (black). The corresponding values of
         $g^2/(\omega_0 \Gamma)$ are $1/6$, $1/4$,
         and $1/3$, respectively. Inset: A fit of
         the $g/\Gamma = 0.324$ curve to the
         functional form of Eq.~(\ref{n_b-fit}) using
         the fitting range $9 \leq \omega_0 t \leq 95$.
         The two curves practically coincide above
         $\omega_0 t = 8$.}
\label{Fig:Occ-Q1-n=0}
\end{figure}

\begin{figure}[t]
\centerline{
\includegraphics[width=80mm]{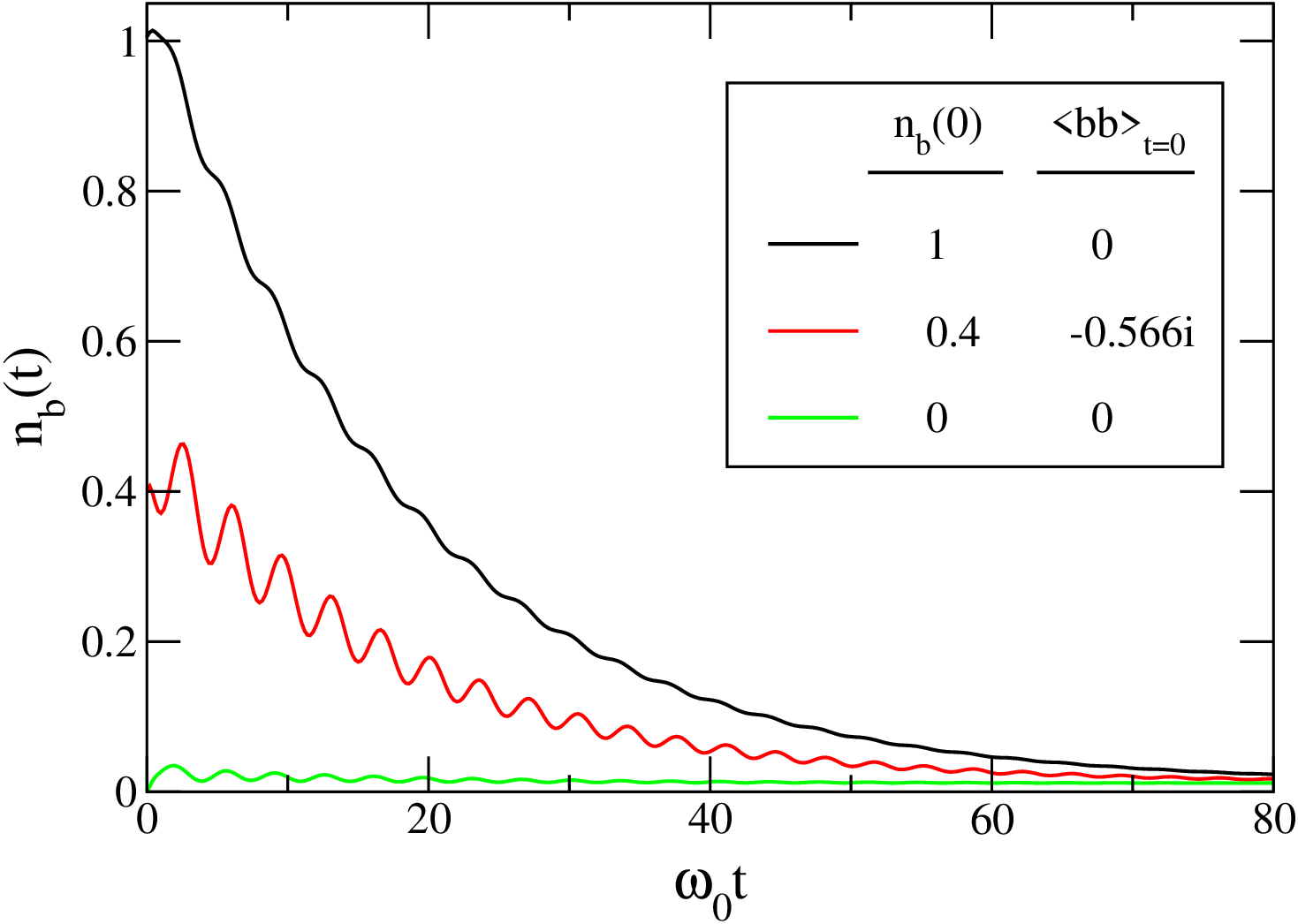}
}\vspace{0pt}
\caption{(Color online)
         Time evolution of the phonon occupancy
         $n_b(t)$ following an abrupt switching on
         of the electron-phonon coupling $g$, for
         $g/\Gamma = 0.324$, $\omega_0/D_d = 0.2$,
         and different initial phononic states.
         Each of the curves with $n_b(0) > 0$
         corresponds to a family of initial
         states whose values of $n_b(0)$ and
         $\langle b b \rangle_{t = 0}$ are specified
         in the legends. Representative states for
         each category are
         $[2 |0 \rangle - i |2 \rangle]/\sqrt{5}$
         (red) and $|1 \rangle$ (green). The curve with
         $n_b(0) = 0$ (black) corresponds exclusively
         to the initial state $|0 \rangle$.}
\label{Fig:Occ-Q1-vs-n}
\end{figure}

Figures~\ref{Fig:Occ-Q1-n=0} and \ref{Fig:Occ-Q1-vs-n}
summarize the time evolution of $n_b(t)$, for different
coupling constants and different initial conditions. In
Fig.~\ref{Fig:Occ-Q1-n=0} we have plotted $n_b(t)$ in
response to an abrupt switching on of the electron-phonon
coupling $g$, with the phonon initially occupying the
empty state $| 0\rangle$ at time $t = 0$. Different
values of $g$ are depicted. Starting from $n_b(0) = 0$,
the time-dependent occupancy first overshoots its new
equilibrium value, to which it then decays with
superimposed oscillations. The oscillatory decay is
well described by the long-time behaviors of $I_1(t)$,
$I_2(t)$, and the term involving the sum over $k$ in the
expression for $F(z,t)$. Indeed, based on our previous
analysis one expects $n_b(t \gg 1/D_d)$ to follow the
functional form
\begin{equation}
n_b(t) = \left [
                 A \sin(2\Omega t + \phi) + B
         \right ]
         e^{-2 t/\tau_{0}} + C ,
\label{n_b-fit}
\end{equation}
with $\Omega$ and $\tau_0$ equal to $\omega$ and $\tau$
of Eqs.~(\ref{omega}) and (\ref{tau}). The inset of
Fig.~(\ref{Fig:Occ-Q1-n=0}) shows a typical fit of the
$g/\Gamma = 0.324$ curve to the functional form of
Eq.~(\ref{n_b-fit}) using the fitting range
$9 \leq \omega_0 t \leq 95$. While some deviations
are seen at shorter times, the two curves are hardly
distinguishable above $\omega_0 t = 8$. Moreover,
the extracted values of $\Omega/\omega0 = 0.896$
and $\tau_0 \omega_0 = 35.5$ fall within 1.2\%
from those of $\omega$ and $\tau$ quoted above.
The agreement between the predicted and extracted
parameters is equally good for the two curves with
the smaller values of $g$, confirming our analytic
predictions for the long-time behavior of $n_b(t)$.

Figure~\ref{Fig:Occ-Q1-vs-n} displays the complementary
dependence of $n_b(t)$ on the initial state of the
localized phonon. As emphasized above, $n_b(t)$ depends
on the initial state of the phonon via two parameters
only: $n_b(0)$ and $\langle bb \rangle_{t=0}$. Hence
each curve with $n_b(0) > 0$ corresponds to a family
of initial states. It is nevertheless useful to have
a particular initial state in mind by assigning a
representative state to each combination of $n_b(0)$
and $\langle bb \rangle_{t=0}$. One possible choice
of states for the two curves with $n_b(0) > 0$ are
$[2 |0 \rangle - i |2 \rangle]/\sqrt{5}$ (red line)
and $|1 \rangle$ (green line). The curve with
$n_b(0) = 0$ (black line) corresponds exclusively
to $|0 \rangle$ as the initial state.

As in Fig.~\ref{Fig:Occ-Q1-n=0}, all curves in
Fig.~\ref{Fig:Occ-Q1-vs-n} can be fit equally well
to the functional form of Eq.~(\ref{n_b-fit})
using the same pair of values for $\Omega$ and
$\tau_0$ that were extracted for $n_b(0) = 0$.
Generally speaking, the larger is $n_b(0)$ the more
pronounced is the component of the pure exponential
decay, while the magnitude of the superimposed
oscillations is more sensitive to
$\langle bb \rangle_{t=0}$.

Another quantity of interest is the time evolution
of the phonon displacement, $Q(t)$. Since $Q$ is
strictly zero for $\epsilon_d = 0$ in thermal
equilibrium, its time evolution remains pinned
to zero unless either $\epsilon_d$ or
$\langle b \rangle_{t=0}$ is nonzero. In this
section we consider the latter possibility
where $\langle b \rangle_{t=0}$ is nonzero. A
straightforward evaluation of $Q(t)$ using
Eq.~(\ref{b_time_dep}) and its Hermitian
conjugate yields
\begin{equation}
Q(t) = \sqrt{2}\,
       {\rm Re} \bigl \{
                         [ I_1(t) + I_2^{\ast}(t) ]
                         \langle b \rangle_{t = 0}
                \bigr \} ,
\end{equation}
whose dependence on the initial state is reduced
to the sole parameter $\langle b \rangle_{t = 0}$.
Writing the latter in terms of its magnitude and phase,
$\langle b \rangle_{t = 0} = |\langle b \rangle|
e^{i\varphi}$, the time-dependent displacement
depends linearly on $|\langle b \rangle|$. The
dependence on $\varphi$ is less transparent as
it requires detailed knowledge of $I_1(t)$ and
$I_2(t)$. Numerical calculations reveal, however,
that the dependence on $\varphi$ is rather weak,
hence we focus our attention hereafter on
$\varphi = 0$.

\begin{figure}[t]
\centerline{
\includegraphics[width=80mm]{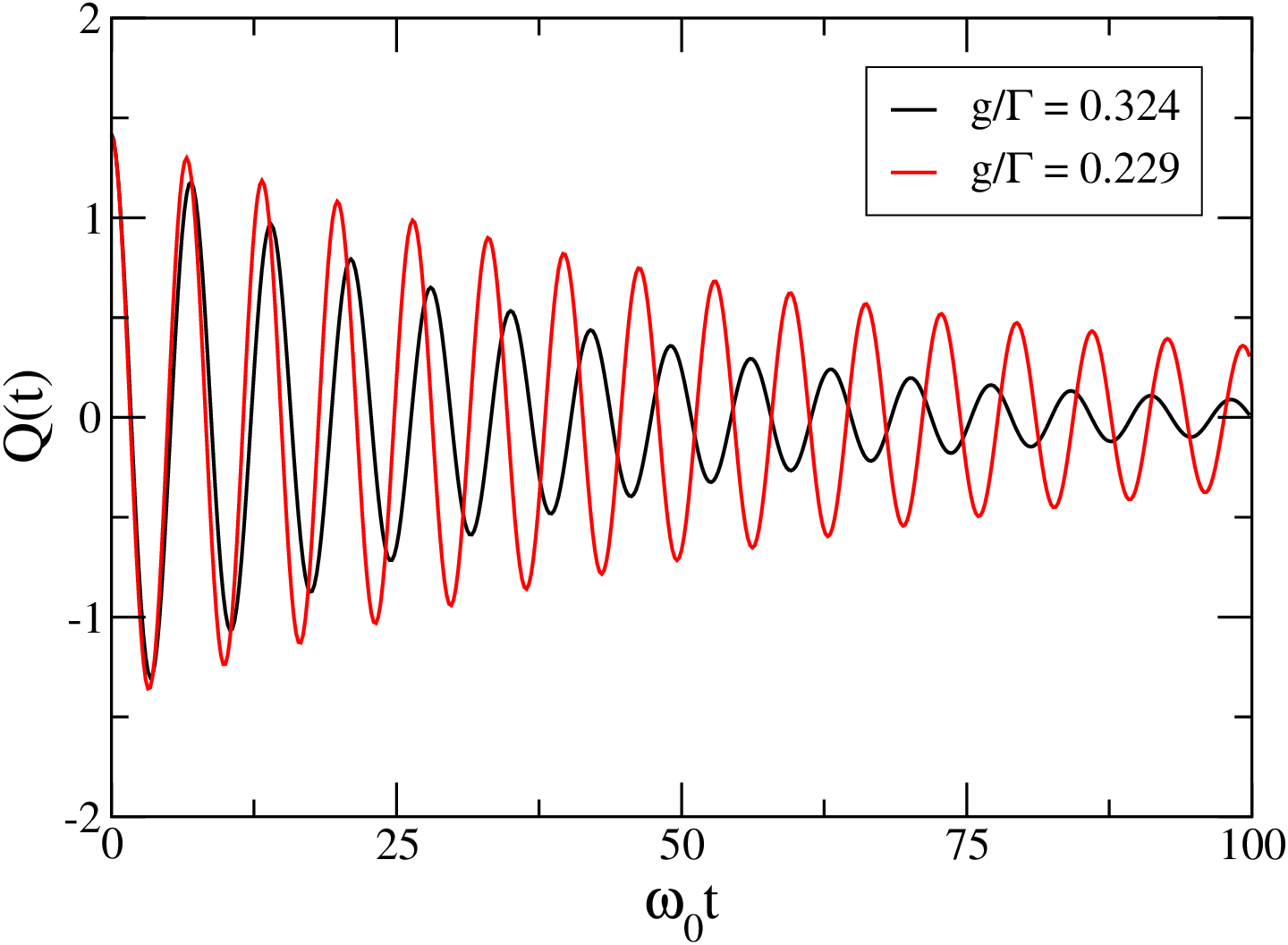}
}\vspace{0pt}
\caption{(Color online)
         Time evolution of the phonon displacement
         $Q(t)$, starting from an initial phonon
         state where $\langle b \rangle_{t=0} = 1$.
         Here $\omega_0/D_d$ equals $0.2$. Two
         representative values of the
         electron-phonon coupling are depicted:
         $g/\Gamma = 0.324$ (black) and
         $g/\Gamma = 0.229$ (red), corresponding
         to $g^2/(\omega_0 \Gamma) = 1/3$ and
         $1/6$, respectively.}
\label{Fig:Q-vs-t-Q1}
\end{figure}

Figure~\ref{Fig:Q-vs-t-Q1} depicts the time evolution
of $Q(t)$ for $\langle b \rangle_{t = 0} = 1$ and
two representative values of the electron-phonon
coupling $g$. As can be seen, $Q(t)$ displays
damped oscillations with a frequency and decay time
that depend on the magnitude of $g$. Indeed, based
on our previous analysis of $I_1(t)$ and $I_2(t)$
one expects the long-time behavior of $Q(t)$ to
follow the functional form
\begin{equation}
Q(t) = A \sin(\Omega t + \phi) e^{-t/\tau_0}
\label{Q-fit}
\end{equation}
with $\Omega$ and $\tau_0$ equal to $\omega$ and
$\tau$, respectively. Fits to Eq.~(\ref{Q-fit})
yield excellent agreement, with values of $\Omega$
and $\tau_0$ that coincide to within less than 1\% with
those extracted from Fig.~\ref{Fig:Occ-Q1-n=0} using
fits to Eq.~(\ref{n_b-fit}).~\cite{comment-on-fits}
Thus, $Q(t)$ displays a relaxation time twice as
long as that of $n_b(t)$ and half the frequency of
oscillations. Such a relation is quite natural for
a classical oscillator where $n_b(t) =
\langle b^{\dagger}(t) \rangle \langle b(t) \rangle
\sim Q^2(t)$, but is less obvious for the
quantum case considered here.

\subsection{Phononic wave function}

Lastly we shall address the time evolution of the
phononic wave function, defined as
\begin{equation}
\left| \psi_{\rm ph}(x,t) \right|^2 =
       \langle
               \delta (\hat{Q}(t) - x)
       \rangle .
\label{phonon-wave-func-def}
\end{equation}
Here, $\hat{Q}(t) = U^{\dagger}(t, 0) \hat{Q} U(t, 0)$
is the phonon displacement operator in its Heisenberg
representation, $x$ is a dimensionless position
coordinate, and averaging is taken with respect to
the initial state of the system. In thermal
equilibrium $\left| \psi_{\rm ph}(x,t) \right|^2$
was calculated by D\'{o}ra,~\cite{Dora_2007} who
showed that it takes a simple Gaussian form. Below
we extend the calculation to nonequilibrium quench
dynamics, allowing for an arbitrary initial phonon
state.

Following D\'{o}ra we begin by rewriting
Eq.~(\ref{phonon-wave-func-def}) as
\begin{equation}
\left| \psi_{\rm ph}(x,t) \right|^2 =
       \int_{-\infty}^{\infty}
            \frac{ds}{2\pi}
            \langle e^{is( \hat{Q}(t) - x )} \rangle .
\end{equation}
Since $\hat{Q}(t)$ is linear in bosonic operators, and
since averaging on the right-hand side is taken with
respect to a product state of the filled Fermi sea
and the initial phonon state,
$\langle e^{is( \hat{Q}(t) - x )} \rangle$ can be
recast as the product of two independent averages of
the conduction-electron and local-phonon components of
$\hat{Q}(t)$. Explicitly, denoting the two components
of $\hat{Q}(t)$ by $\hat{Q}_c(t)$ and $\hat{Q}_b(t)$
one has that
\begin{equation}
\langle e^{is \hat{Q}(t)} \rangle
    = \langle e^{is \hat{Q}_b(t)} \rangle_{\rm b}
      \langle e^{is \hat{Q}_c(t)} \rangle_{\rm FS} \, ,
\label{ave-Qc-Qb}
\end{equation}
where $\langle \ldots \rangle_{\rm FS}$ and
$\langle \ldots \rangle_{\rm b}$ stand for averaging
with respect to the filled Fermi sea and the initial
phononic state, respectively.

Each of the two averages in Eq.~(\ref{ave-Qc-Qb})
can be evaluated in turn using standard bosonic
techniques. Consider first the conduction-electron
component. As the average is taken with respect to
the ground state of a free bosonic bath one can use
the identity $\langle e^{\hat{A}} \rangle =
e^{\langle \hat{A}^2 \rangle/2}$, applicable to
any operator $\hat{A}$ that is linear in bosonic
creation and annihilation operators. This results in
\begin{equation}
\langle e^{is \hat{Q}_c(t)} \rangle_{\rm FS} =
    \exp \left [
                 - \frac{s^2}{2}
                   \langle
                           \hat{Q}^2_c(t)
                   \rangle_{\rm FS}
         \right ] .
\label{ave-Qc}
\end{equation}
Moving on to the local phonon component, we first note
that $\hat{Q}_b(t)$ has the explicit form
\begin{equation}
\hat{Q}_b(t) = I_3(t) b^{\dagger}
               + I_3^{*}(t) b ,
\end{equation}
where
\begin{eqnarray}
I_3(t) &=& \frac{1}{\sqrt{2}}
           \bigl [
                   I_1^{*}(t) + I_2(t)
           \bigr ]
        = \sqrt{2} \omega_0 \lambda^2
          \sum_{k>0} \xi_k^2
                     |g(\epsilon_k+i\eta)|^2
\nonumber \\
       && \times
                     \left[
                            (\epsilon_k - \omega_0)
                             e^{i\epsilon_k t} +
                            (\epsilon_k + \omega_0)
                             e^{-i\epsilon_k t}
                     \right] .
\end{eqnarray}
Next we use the identity $e^{\hat{A} + \hat{B}}
= e^{\hat{A}} e^{\hat{B}} e^{[\hat{B}, \hat{A}]/2}$,
applicable to any two operators $\hat{A}$ and
$\hat{B}$ whose commutator is a $c$-number, to
write
\begin{equation}
\langle e^{is \hat{Q}_b(t)} \rangle_{\rm b} =
        e^{-(s^2/2)|I_3(t)|^2}
        \langle
                e^{i s I_3(t) b^{\dagger}}
                e^{i s I^*_3(t)b}
        \rangle_{\rm b} .
\label{ave-Qb}
\end{equation}
The combination of Eqs.~(\ref{ave-Qc-Qb}),
(\ref{ave-Qc}), and (\ref{ave-Qb}) then yields
\begin{equation}
|\psi_{\rm ph}(x,t)|^2 =
      \int_{-\infty}^{\infty}
           \frac{ds}{2\pi}
           e^{-\gamma(t)s^2 - isx}
           \langle
                   e^{i s I_3(t) b^{\dagger}}
                   e^{i s I^*_3(t) b}
           \rangle_{\rm b}
\label{phonon-wave-function}
\end{equation}
with
\begin{equation}
\gamma(t) = \omega_0^2 \lambda^2
            \sum_{k > 0}
                  \xi_k^2
                  \left |
                          K(\epsilon_k - i\eta, t)
                  \right |^2
            + \frac{1}{2}|I_3(t)|^2
\end{equation}
and
\begin{eqnarray}
K(z, t) &=& g(z) e^{i z t} + 2 \omega_0 \lambda^2
            \sum_{q > 0} \xi_q^2
            |g(\epsilon_q + i\eta)|^2
\nonumber \\
       &&
            \times
            \left(
                   \frac{e^{i \epsilon_q t}}
                        {\epsilon_ q -z}
                   + \frac{e^{-i \epsilon_q t}}
                          {\epsilon_q + z}
            \right) .
\end{eqnarray}

Equation~(\ref{phonon-wave-function}) allows one
to calculate the phononic wave function for
arbitrary time $t > 0$ and any initial phononic state.
Before turning to concrete examples let us address
some generic features of $|\psi_{\rm ph}(x, t)|^2$.
Since $I_3(t)$ decays to zero as $t \to \infty$,
the expectation value on the right-hand side of
Eq.~(\ref{phonon-wave-function}) reduces
asymptotically to one regardless of the initial
state of the phonon. Furthermore, repeating the
same type of analysis as beforehand one finds
that the term involving the sum over $q$ in the
expression for $K(\epsilon_k - i\eta, t)$ decays
to zero with the relaxation time $\tau$, and
that $\gamma(t)$ decays to its asymptotic value
\begin{equation}
\gamma = \lim_{t \to \infty} \gamma(t)
       = \lambda^2 \omega_0^2
         \sum_{k > 0} \xi_k^2
                \left|
                        g(\epsilon_k + i\eta)
                \right|^2
\end{equation}
with the reduced relaxation time $\tau/2$. The
phononic wave function thus takes the asymptotic
Gaussian form
\begin{equation}
|\psi_{\rm ph}(x)|^2
     = \lim_{t \to \infty}|\psi_{\rm ph}(x,t)|^2
     = \frac{1}{2\sqrt{\pi \gamma}}
       \exp \left [
                    -\frac{x^2}{4 \gamma}
            \right ] .
\label{Asymptotic-WF}
\end{equation}
Lastly, recognizing that $\gamma$ is half the
thermalized expectation value of $\hat{Q}^2$, i.e.,
$2 \gamma = \langle \hat{Q}^2 \rangle_{\rm eq}$,
we recover hereby the equilibrium result of
D\'{o}ra~\cite{Dora_2007} at long times.

While the asymptotic form of the phononic wave
function is independent of the initial state of the
phonon, the associated relaxation time does depend on
whether $\langle b \rangle_{t = 0}$ is zero or not.
To see this we note that $|\psi_{\rm ph}(x,t)|^2$
has two sources of time dependence originating
from $\gamma(t)$ and $I_3(t)$. While $\gamma(t)$
decays to its asymptotic value $\gamma$ with a
relaxation time equal to $\tau/2$, $I_3(t)$ decays
to zero with a relaxation time that is twice as long.
The relaxation of $|\psi_{\rm ph}(x,t)|^2$ depends
then on whether the expectation value on the
right-hand side of Eq.~(\ref{phonon-wave-function})
has a contribution that is linear in $I_3(t)$
and $I^{\ast}_3(t)$ or not. If
$\langle b \rangle_{t = 0} = 0$ there is no
such linear contribution, hence
$|\psi_{\rm ph}(x,t)|^2$ approaches its asymptotic
form with the relaxation time $\tau/2$. If, on
the other hand, $\langle b \rangle_{t = 0}$ is
nonzero then there is such a contribution and
$|\psi_{\rm ph}(x,t)|^2$ relaxes on a longer
time scale equal to $\tau$.

\begin{figure}[t]
\centerline{
\includegraphics[width=80mm]{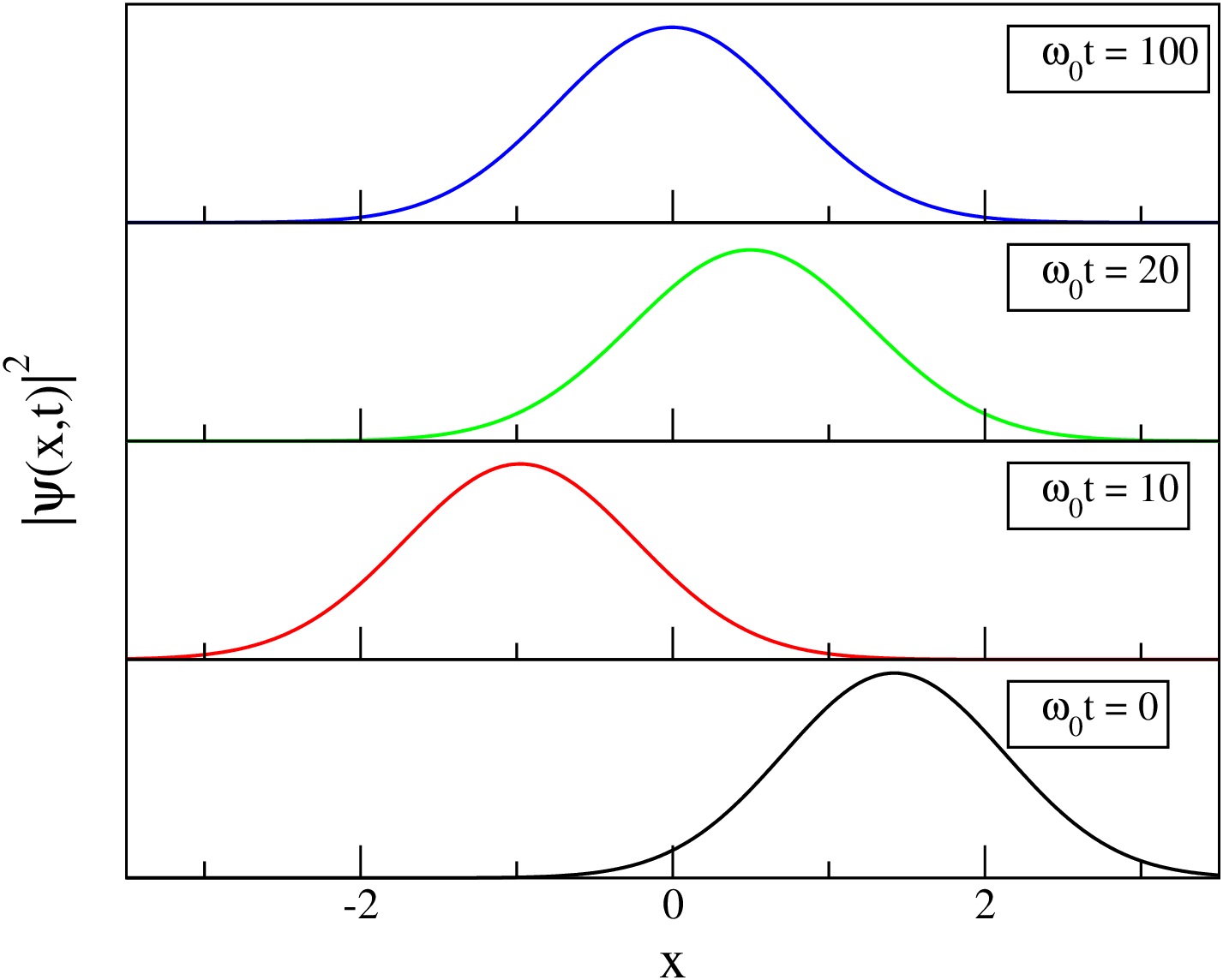}
}\vspace{0pt}
\caption{(Color online)
         Time evolution of the phononic wave function
         $|\psi(x, t)|^2$, starting from an initial
         coherent state with $\lambda = 1$. Here
         $\omega_0/D_d = 0.2$ and
         $g/\Gamma = 0.324$.}
\label{Fig:psi-Q1-b=1}
\end{figure}

Although Eq.~(\ref{phonon-wave-function}) applies
to any initial phonon state, of particular interest
are those cases where the phonon initially occupies
either a coherent state or an eigenstate of
$\hat{n}_b = b^{\dagger} b$. If the initial state
is a coherent state, i.e., $b |\psi_0 \rangle =
\lambda |\psi_0 \rangle$, then
\begin{equation}
\langle
        e^{i s I_3(t) b^{\dagger}}
        e^{i s I^*_3(t) b}
\rangle_{\rm b}
= e^{ i s 2 {\rm Re} \{ I^*_3(t) \lambda \} }
= e^{i s Q(t)} ,
\end{equation}
resulting in
\begin{equation}
|\psi_{\rm ph}(x,t)|^2 =
       \frac{1}{2\sqrt{\pi \gamma(t)}}
       \exp \left [
                    -\frac{(x - Q(t))^2}
                          {4 \gamma(t)}
            \right ] .
\end{equation}
The phononic wave function is therefore a simple
Gaussian, characterized by the time-dependent
average $Q(t)$ and the time-dependent width
$\sigma(t) = \sqrt{2 \gamma(t)}$. If the initial
state is an eigenstate of $\hat{n}_b$ with the
eigenvalue $n$ the phononic wave function is
somewhat more convoluted, given by the formal
expression
\begin{equation}
| \psi_{\rm ph}(x,t) |^2 =
        \sum_{m = 0}^n
              \left(
              \!\!
                     \begin{array}{c}
                            n \\ m
                     \end{array}
              \!\!
              \right)
              \frac{|I_3(t)|^{2m}}
                   {2 m! \sqrt{\pi\gamma(t)}}
              \frac{d^{2m}\;}
                   {d x^{2m}}
                   e^{-x^2/4\gamma(t)} .
\label{wave-func-n}
\end{equation}
Alternatively, Eq.~(\ref{wave-func-n}) can be rewritten
using Hermite polynomials as
\begin{equation}
| \psi_{\rm ph}(x,t) |^2 =
        \sum_{m = 0}^n
              \left(
              \!\!
                     \begin{array}{c}
                            n \\ m
                     \end{array}
              \!\!
              \right)
              \frac{|I_3(t)|^{2m}}
                   {m! \sqrt{\pi}
                    [4 \gamma(t)]^{m + 1/2}}
                   H_n(y) e^{-y^2} ,
\label{wave-func-n-Hermite}
\end{equation}
with $y = x/\sqrt{4\gamma(t)}$.

\begin{figure}[t]
\centerline{
\includegraphics[width=80mm]{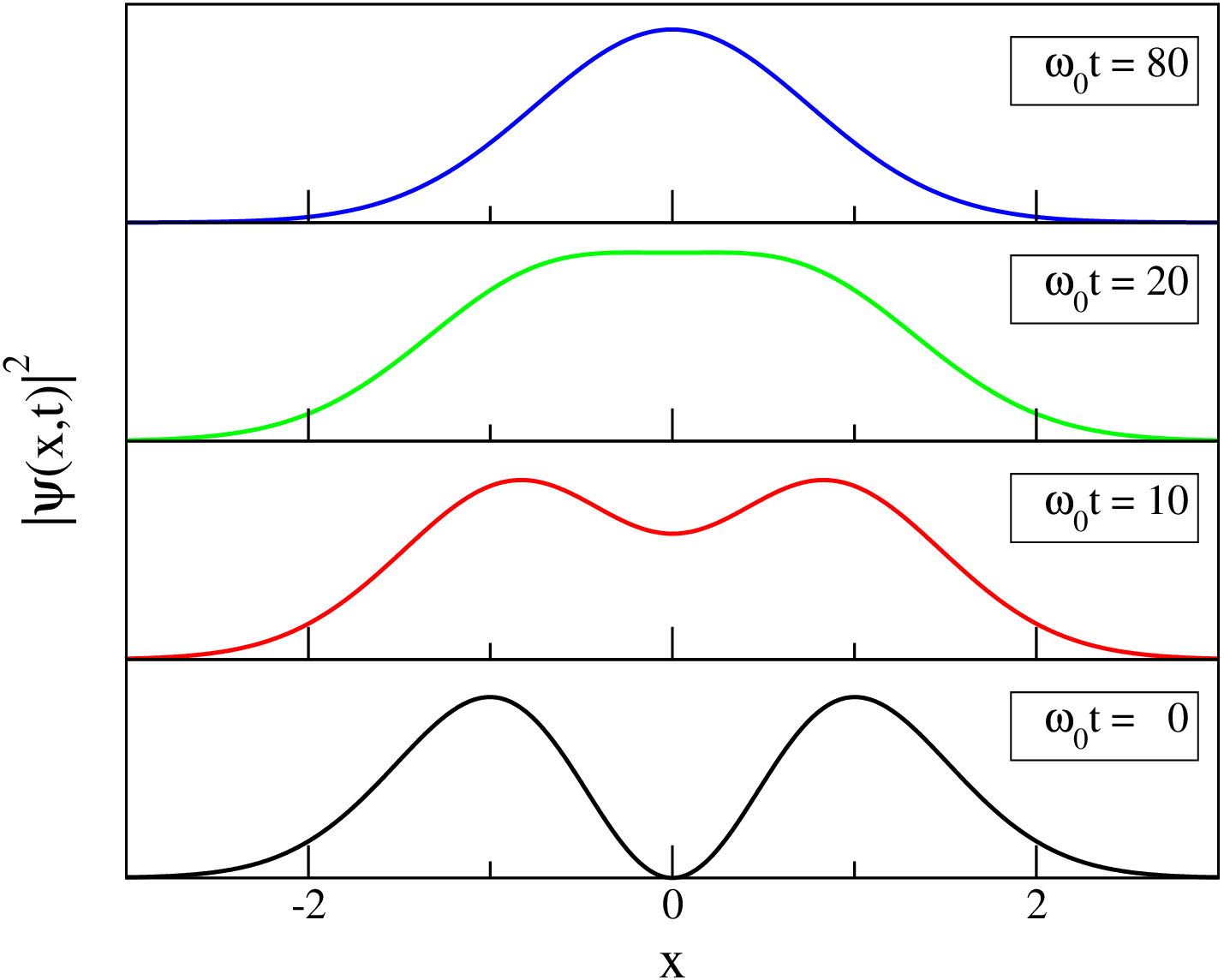}
}\vspace{0pt}
\caption{(Color online)
         Same as Fig.~\ref{Fig:psi-Q1-b=1}, starting
         from an initial phonon state where
         $\hat{n}_b = 1$.}
\label{Fig:psi-Q1-n=1}
\end{figure}
 
The time evolution of the phononic wave function
is displayed in Figs.~\ref{Fig:psi-Q1-b=1} and
\ref{Fig:psi-Q1-n=1} for two representative initial
configurations of the phonon: a coherent state with
$\lambda = 1$ (Fig.~\ref{Fig:psi-Q1-b=1}) and an
eigenstate of $\hat{n}_b$ with the eigenvalue $n = 1$
(Fig.~\ref{Fig:psi-Q1-n=1}). Starting from a coherent
state, the phononic wave function evolves through
a sequence of Gaussians, as can be seen in
Fig.~\ref{Fig:psi-Q1-b=1}. The center of the
Gaussian, $Q(t)$, oscillates from $\sqrt{2}$ at time
$t = 0$ to zero as $t \to \infty$ according to the
black curve in Fig.~\ref{Fig:Q-vs-t-Q1}, while its
width increases form $1/\sqrt{2}$ to $1$. In contrast,
the phononic wave function undergoes a qualitative
change in shape when starting from $\hat{n}_d = 1$.
Here $|\psi(x, t)|^2$ is initially composed of two
symmetric peaks that gradually merge to a single
Gaussian at long times. This behavior can be
understood from the explicit form of
Eq.~(\ref{wave-func-n-Hermite}) with $n = 1$:
\begin{equation}
|\psi_{\rm ph}(x,t)|^2 =
       \left [
               1 - \frac{|I_3(t)|^2}{2 \gamma(t)}
                 + \frac{|I_3(t)|^2}{4 \gamma(t)^2} x^2
       \right ]
       \frac{e^{-x^2/4\gamma(t)}}
            {2 \sqrt{\pi \gamma(t)} } .
\end{equation}
At $t = 0$ one can show that $|I_3(0)|^2 = 2 \gamma(0)
= 1/2$, resulting in
\begin{equation}
|\psi_{\rm ph}(x, t = 0)|^2 =
       \frac{2}{\sqrt{\pi}} x^2 e^{-x^2} .
\end{equation}
As time increases $|I_3(t)|^2$ gradually decays to
zero, leaving us with the thermalized Gaussian of
Eq.~(\ref{Asymptotic-WF}). The transition between the
two forms of the wave function is therefore driven
by the relaxation of $|I_3(t)|^2$, which happens on
a time scale of $\tau/2$.

\subsection{Thermalization in the presence of integrability}

Tracking the time evolution of the phononic occupancy,
displacement, and wave function in response to
switching $g$ on, we observed in the previous
subsections that all quantities eventually approach
their thermal equilibrium values with respect to
the full Hamiltonian. In other words, the system
thermalizes at long times. Indeed, this limit was
rigorously shown by Ambegaokar~\cite{Ambegaokar_2007}
for a class of bosonic models that include our
Hamiltonian of interest. More generally, it was
shown by Doyon and Andrei~\cite{Doyon-Andrei} in
the context of interacting quantum dots that
\begin{equation}
\lim_{t \to \infty}
      U^{\dagger}(t,0) e^{-\beta {\cal H}_0} U(t,0)
      = e^{-\beta {\cal H}} ,
\end{equation}
provided the bath is a large Fermi sea and ${\cal H}
- {\cal H}_0$ is a local perturbation, as is the case
here. However, one may wonder at this point how can an
integrable system thermalize given the infinite set of
conservation laws it possesses [specifically, the
occupation numbers $\alpha^{\dagger}_k \alpha_k$,
see Eq.~(\ref{eq_ham_in_diag_basis})]?

To address this question consider the quench dynamics
starting from an excited Fermi sea
\begin{equation}
| n_{k_1}, \ldots, n_{k_N} \rangle_{g = 0}
      = \left (
                a^{\dagger}_{k_1}
        \right)^{n_{k_1}} \cdots
        \left(
                a^{\dagger}_{k_N}
        \right)^{n_{k_N}} |0\rangle_{g = 0} ,
\end{equation}
obtained by creating several particle-hole excitations
with momenta $k_1, k_2, \ldots, k_N$ above the filled
Fermi sea of the unperturbed system. Starting from
the product state $|\psi_0 \rangle = | n_b \rangle
\otimes | n_{k_1}, \ldots, n_{k_N} \rangle_{g = 0}$
of the excited Fermi sea and a local phonon state
with the occupation number $n_b$, one can repeat
the calculation of Sec.~\ref{sec:Q1-nb} to track
the time evolution of the phonon occupancy. At
long time one finds
\begin{eqnarray}
n_b(t \to \infty) &=& \lambda^2
      \sum_{q > 0}
           \xi_q^2 \left|
                          g(\epsilon_q + i\eta)
                   \right|^2
                   (\epsilon_q - \omega_0)^2
\nonumber \\
  &+& 2\lambda^2
       \sum_{j = 1}^{N}
            \xi_{k_j}^2 \left|
                               g(\epsilon_{k_j} + i\eta)
                        \right|^2
                        \left(
                               \epsilon_{k_j}^2
                               + \omega_0^2
                        \right) n_{k_j}
\nonumber \\
  &=& _{g}\langle n_{k_1}, \ldots, n_{k_N}
                \left|
                       b^{\dagger}b
                \right|
         n_{k_1}, \ldots, n_{k_N} \rangle_{g} \, ,
\end{eqnarray}
where
\begin{equation}
| n_{k_1}, \ldots, n_{k_N} \rangle_{g}
      = \left (
                \alpha^{\dagger}_{k_1}
        \right)^{n_{k_1}} \cdots
        \left(
                \alpha^{\dagger}_{k_N}
        \right)^{n_{k_N}} |0\rangle_{g}
\end{equation}
is the corresponding eigenstate of the full Hamiltonian,
obtained by creating scattering-state excitations with
identical quantum numbers above the ground state of the
full system. Thus, while the initial state of the local
phonon is wiped out in the course of the evolution, the
quantum numbers characterizing the initial state of
the Fermi sea are preserved. In other terms, the
conservation laws constrain the bulk but not the local
degrees of freedom. Since $\xi^2_{k_j}$ scales as
$1/L$, local observables are independent of the initial
state of the Fermi sea as long as the initial excitation
energy is not extensive, i.e.,
\begin{equation}
\frac{1}{L} \sum_{j = 1}^{N} n_{k_j} \to 0
\end{equation}
in the thermodynamic limit.
We therefore conclude that the local phonon thermalizes
while the bath does not, and that the conservation laws
which constrain the bath dynamics do allow for a
generic evolution of local degrees of freedom.

\section{Abrupt Change of Phonon Frequency}
\label{sec:frequency-quench}

The second quench dynamics we consider is the
response to a sudden change in the phonon
frequency. Namely, the system is taken to
occupy the ground state of the Hamiltonian of
Eq.~(\ref{H_bosonic}) at time $t = 0$ when the
phonon frequency is abruptly shifted from $\omega_0$
to $\omega_1 = \omega_0 + \delta \omega > 0$. In
contrast to the electron-phonon coupling, which
is difficult to control in actual devices, the
frequency of vibrations can be tuned electrically
in suspended carbon nanotubes.~\cite{Frequency-tuning}
This offers a potential realization of the
present scenario. For concreteness we restrict
attention in this section to $\epsilon_d = 0$
and zero temperature, though both restrictions
can be relaxed. Accordingly, our interest will
center on $n_b(t)$, as $Q(t)$ is pinned by
symmetry to zero. Similarly, the phononic wave
function retains a Gaussian form centered about
$x = 0$ at arbitrary time $t$, with a
time-dependent width equal to
$\sqrt{\langle \hat{Q}^2(t) \rangle}$.

\subsection{Time evolution of phononic operators}

The general strategy for calculating the time evolution
of physical observables is similar to the one taken
in the previous section, except that the initial
Hamiltonian is now the full Hamiltonian ${\cal H}$
of Eq.~(\ref{H_bosonic}) with $\tilde{\epsilon}_d$
set to zero, and the final Hamiltonian is given by
${\cal H}' = {\cal H} + \delta {\cal H}$ with
\begin{equation}
\delta {\cal H} = \delta \omega b^{\dagger} b .
\end{equation}
The technical details are slightly more cumbersome,
though, since the time evolution of $b^{\dagger}(t)$
is carried out by expanding $b^{\dagger}$ in terms
of the eigenmodes $\gamma_k$ and $\gamma^{\dagger}_k$
of the final Hamiltonian ${\cal H}'$, whereas the
evaluation of expectation values requires an expansion
of $b^{\dagger}(t)$ in terms of the eigenmodes
$\alpha_k$ and $\alpha^{\dagger}_k$ of the initial
Hamiltonian ${\cal H}$. In other words, one needs
to know how to convert from the eigenmodes of
${\cal H}'$ to those of ${\cal H}$.

There are two approaches one can take to achieve this
goal. The first is to invert Eq.~(\ref{eq_sc_state})
and its Hermitian conjugate in order to express $a_q$,
$a^{\dagger}_q$, $b$ and $b^{\dagger}$ in terms of
the eigenmodes of ${\cal H}$, and to plug the resulting
expressions into the expansion of $\gamma^{\dagger}_k$
in terms of $a_q$, $a^{\dagger}_q$, $b$ and $b^{\dagger}$.
An alternative approach is to directly express
$\gamma^{\dagger}_k$ in terms of $\alpha_q$ and
$\alpha^{\dagger}_q$ by solving the modified
Lippmann-Schwinger equation
\begin{equation}
[\gamma^{\dagger}_k, {\cal H}']
     = -\epsilon_k \gamma^{\dagger}_k
     + i\eta (\alpha^{\dagger}_k - \gamma^{\dagger}_k) .
\end{equation}
To this end, it is necessary to first write ${\cal H}'$
in terms of the eigenmodes of ${\cal H}$, which
follows directly from Eqs.~(\ref{eq_ham_in_diag_basis})
and (\ref{b_via_sc_state}). As we prove in
Appendix~\ref{app_scat_state}, the two methods of
computation are equivalent, allowing us to use the
latter approach which is more concise. Differing
all details of the calculation to the Appendix we
quote here only the end result:
\begin{eqnarray}
\gamma^{\dagger}_k &=& \alpha^{\dagger}_k
       + 2 \delta \omega \lambda^2 \xi_k
           \tilde{g}(\epsilon_k + i\eta)
\nonumber \\
       && \;\;\;\;\;\;\;\;\,
          \times\!
        \sum_{q > 0} \xi_q
        \bigg [
                 g(\epsilon_q - i\eta)
                 \frac{\epsilon_k \epsilon_q
                       + \omega_0 \omega_1}
                      {\epsilon_k - \epsilon_q+i\eta}
                 \alpha^{\dagger}_q
\nonumber \\
       && \;\;\;\;\;\;\;\;\,
                - g(\epsilon_q + i\eta)
                  \frac{\epsilon_k \epsilon_q
                        - \omega_0 \omega_1}
                       {\epsilon_k + \epsilon_q + i\eta}
                 \alpha_q
        \bigg ] ,
\label{gamma_k}
\end{eqnarray}
where
\begin{equation}
\tilde{g}(z) =
      \frac{1}
           {z^2 - \omega_1^2 - 2 \omega_1 \Sigma(z)}
\label{g_tilde}
\end{equation}
is the same function of Eq.~(\ref{eq_g_defined})
with $\omega_0$ replaced by $\omega_1$.

Since ${\cal H}'$ has the same exact form as the
Hamiltonian ${\cal H}$ of Eq.~(\ref{H_bosonic})
only with $\omega_0$ replaced by $\omega_1$, one
can borrow all results derived previously in
Sec.~\ref{Sec_exact_diag} for the latter
Hamiltonian. In particular, ${\cal H}'$ is
diagonal in the new basis set,
\begin{equation}
{\cal H}' = \sum_{k > 0}
            \epsilon_k \gamma^{\dagger}_k \gamma_k ,
\end{equation}
and the expansions of $b^{\dagger}$ and $b^{\dagger}(t)$
detailed in Eqs.~(\ref{b_via_sc_state}) and
(\ref{b_time_dep}) still hold upon substituting
$\omega_1$, $\tilde{g}$, $\gamma_k$, and
$\gamma^{\dagger}_k$ in for $\omega_0$, $g$,
$\alpha_k$, and $\alpha^{\dagger}_k$, respectively.
Plugging Eq.~(\ref{gamma_k}) and its Hermitian
conjugate into Eq.~(\ref{b_time_dep}) we finally
obtain the desired expansion of $b^{\dagger}(t)$
in terms of the $\alpha_k$'s:
\begin{eqnarray}
b^{\dagger}(t) &=& \lambda
       \sum_{k > 0} \xi_k
            \big[
                  \tilde{g}(\epsilon_k - i\eta)
                           (\epsilon_k + \omega_1)
                  e^{i\epsilon_k t}
\nonumber \\
       && \;\;\;\;\;\;\;\;\;
                  + 2 \delta \omega
                    g(\epsilon_k - i\eta)
                    J(\epsilon_k - i\eta,t)
            \big]
            \alpha^{\dagger}_k
\nonumber \\
      &+&
      \lambda \sum_{k > 0} \xi_k
            \big[
                  \tilde{g}(\epsilon_k + i\eta)
                  (-\epsilon_k + \omega_1)
                  e^{-i\epsilon_k t}
\nonumber \\
       && \;\;\;\;\;\;\;\;\;
                  + 2 \delta \omega
                    g(\epsilon_k + i\eta)
                    J(-\epsilon_k - i\eta,t)
            \big]
            \alpha_k ,
\label{b-Q2-via-alpha}
\end{eqnarray}
where we have defined the auxiliary function
\begin{eqnarray}
J(z, t)\!&=&\!\lambda^2\!
           \sum_{k > 0} \xi_k^2
                 |\tilde{g}(\epsilon_k + i\eta)|^2
           \big [
                  (\epsilon_k + \omega_1)
                  \frac{\epsilon_k z + \omega_0 \omega_1}
                       {\epsilon_k - z}
                  e^{i\epsilon_k t}
\nonumber \\
        && \;\;\;\;\;\;\;\;\;\;
                + (\epsilon_k - \omega_1)
                  \frac{\epsilon_k z - \omega_0 \omega_1}
                       {\epsilon_k + z}
                  e^{-i\epsilon_k t}
           \big ] .
\end{eqnarray}

\subsection{Phononic occupancy}

With Eq.~(\ref{b-Q2-via-alpha}) at hand, we are
now in position to evaluate expectation values
pertaining to the local phonon mode $b^{\dagger}$.
Focusing on the time evolution of the phonon occupancy
$n_b(t) = \langle b^{\dagger}(t) b(t) \rangle$, we
note that $b^{\dagger}(t) b(t)$ is quadratic in
$\alpha^{\dagger}_k$ and $\alpha_k$. Since the
expectation value is taken with respect to the ground
state of the initial Hamiltonian ${\cal H}$, the only
nonzero contributions stem from the diagonal terms
$\alpha_k \alpha^{\dagger}_k$, resulting in
\begin{eqnarray}
n_b(t) &=& \lambda^2
           \sum_{k > 0} \xi_k^2
                 | \tilde{g}(\epsilon_k + i\eta)
                   (-\epsilon_k + \omega_1)
                   e^{-i\epsilon_k t}
\nonumber \\
       && \;\;\;\;\;\;\;\;\;\;
                 + 2 \delta \omega
                   g(\epsilon_k + i\eta)
                   J(-\epsilon_k - i\eta,t) |^2 .
\label{n_b-Q2}
\end{eqnarray}

\begin{figure}[t]
\centerline{
\includegraphics[width=80mm]{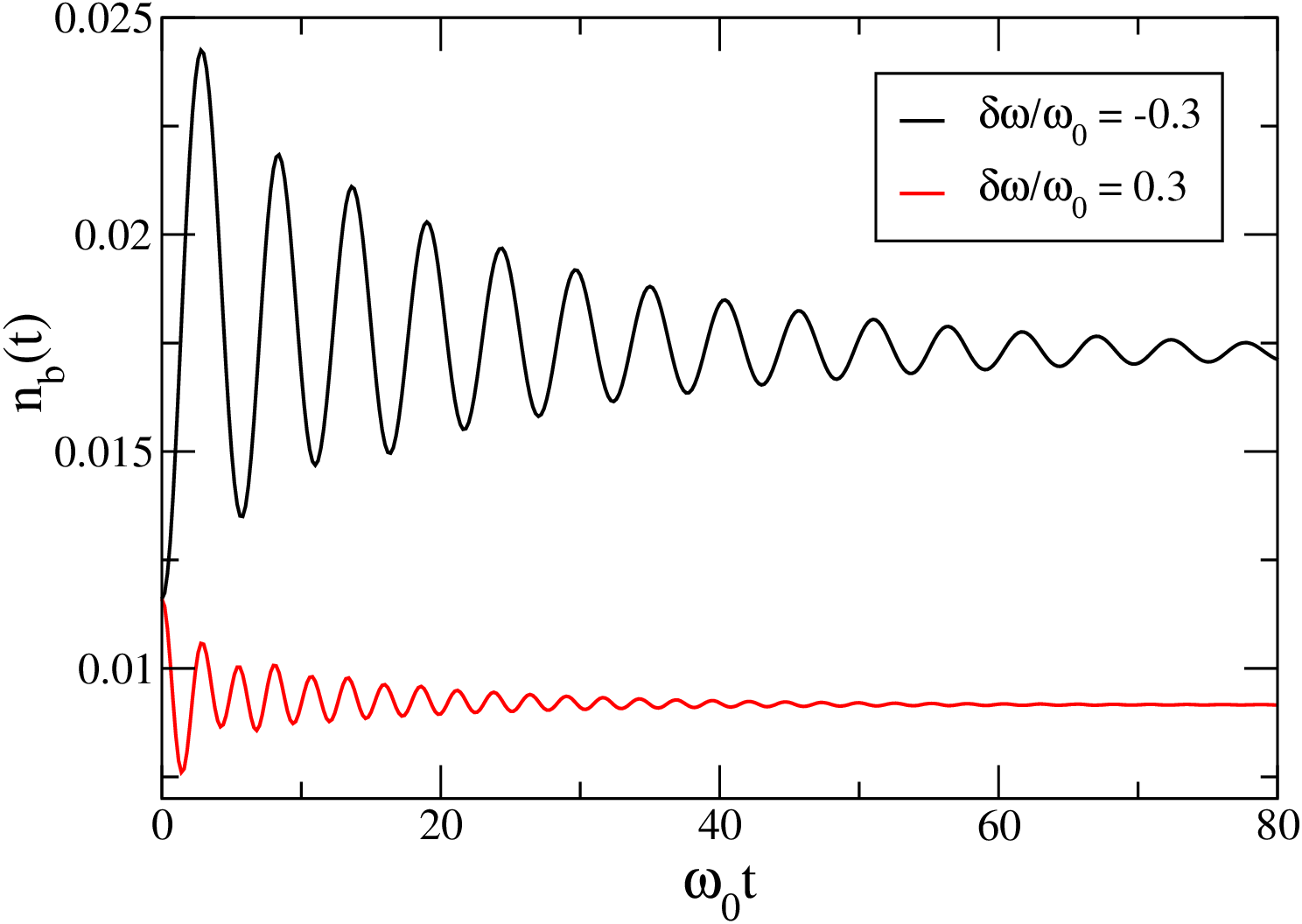}
}\vspace{0pt}
\caption{(Color online)
         Time evolution of the phononic occupancy
         following an abrupt shift in the phonon
         frequency from $\omega_0/D_d = 0.2$ to
         $\omega_1 = \omega_0 + \delta\omega$ with
         $\delta \omega = \pm 0.3\omega_0$. The
         electron-phonon interaction is held fixed
         at $g/\Gamma = 0.324$.}
\label{Fig:Phonon-occupancy-Q2}
\end{figure}

At $t = 0$, Eq.~(\ref{n_b-Q2}) properly reduces to
the equilibrium expectation value of $\hat{n}_b$
with respect to ${\cal H}$ specified in
Eq.~(\ref{eq_occ_infty}). To see this we note that
Eq.~(\ref{b-Q2-via-alpha}) must coincide at time
$t = 0$ with Eq.~(\ref{b_via_sc_state}), as both
expressions offer an expansion of $b^{\dagger}$ in
terms of the $\alpha_k$'s and $\alpha^{\dagger}_k$'s.
Equating the corresponding expansion coefficients
one finds the identity
\begin{eqnarray}
g(\epsilon_k + i\eta)(-\epsilon_k + \omega_0)
          &=&  2 \delta \omega
               g(\epsilon_k + i\eta)
               J(-\epsilon_k - i\eta, 0)
\nonumber \\
        &+& \tilde{g}(\epsilon_k + i\eta)
                      (-\epsilon_k + \omega_1) ,
\end{eqnarray}
from which the equivalence of Eq.~(\ref{eq_occ_infty})
and Eq.~(\ref{n_b-Q2}) at time $t = 0$ immediately
follows. In the opposite limit of long times
Eq.~(\ref{n_b-Q2}) reproduces the new equilibrium
expectation value of $\hat{n}_b$ with respect to
${\cal H}'$. Indeed, using a similar analysis as
beforehand one can show that $J(-\epsilon_k - i\eta,t)$
decays to zero with the same relaxation time $\tau$
and frequency of oscillations $\omega$ as listed
in Eqs.~(\ref{omega}) and (\ref{tau}), subject to
the substitution of $\omega_0$ with $\omega_1$.
This in turn leaves us at long times with
\begin{equation}
n_b(t \to \infty) = \lambda^2
          \sum_{k>0} \xi_k^2
               \left |
                       \tilde{g}(\epsilon_k + i\eta)
               \right|^2
               (\epsilon_k - \omega_1)^2 ,
\end{equation}
which is the thermalized expectation value of
$\hat{n}_b$ with respect to ${\cal H}'$. Thus,
as expected, $n_b(t)$ interpolates between the
two equilibria expectation values of $\hat{n}_b$.

Figure~\ref{Fig:Phonon-occupancy-Q2} shows the
complete time evolution of $n_b(t)$ for two opposite
shifts of the phonon frequency. Once again the
curves take the form of damped oscillations with
the relaxation time $\tau/2$ and the frequency
of oscillations $2 \omega$. Consequently, the
decay time and frequency of oscillations differ
substantially between $\delta \omega = 0.3\omega_0$
and $\delta \omega = -0.3\omega_0$, in accordance
with the substitution $\omega_0 \to \omega_1 =
(1 \pm 0.3) \omega_0$ in Eqs.~(\ref{omega}) and
(\ref{tau}). The larger is $\omega_1$ the smaller
are the new thermalized expectation value of
$\hat{n}_b$ and the amplitude of damped
oscillations that $n_b(t)$ undergoes.

\section{Abrupt shift of energy level}
\label{sec:level-quench}

The third and final quench scenario we consider is
the response to an abrupt shift in the electronic
energy level, which has been held fixed up until now
at resonance with the Fermi energy. Specifically, we
assume that the system resides at time $t < 0$ in
its ground state for $\epsilon_d = 0$, when a nonzero
$\epsilon_d$ is suddenly switched on. This has the
effect of breaking particle-hole symmetry, dynamically
generating a nonzero displacement $Q(t)$ of the local
phonon along with deviations of the level occupancy
from half filling [i.e., $n_d(t) \neq 1/2$]. These two
observables will be our main focus of interest. Of
the different quench scenarios under consideration
the present one is by far the most accessible
experimentally, as the energy level $\epsilon_d$ can
be controlled quite efficiently using suitable gate
voltages.

The foundations for calculating $Q(t)$ and $n_d(t)$
in this scenario have been laid down in
Sec.~\ref{subsec:nonzero-ed}. Specifically, from
Eq.~(\ref{b_zero_via_sc_state}) one has that
\begin{eqnarray}
\tilde{b}^{\dagger}(t) &=& \lambda
                \sum_{k>0} \xi_k
                     \bigg[
                            g(\epsilon_k - i\eta)
                             (\epsilon_k + \omega_0)
                              e^{i \epsilon_k t}
                              \beta^{\dagger}_k
\nonumber \\
               && \;\;\;\;\;\;\;\;\;\;\;\;\;\; -
                            g(\epsilon_k + i\eta)
                             (\epsilon_k-\omega_0)
                              e^{-i \epsilon_k t}
                             \beta_k
                     \bigg]
\end{eqnarray}
which, when combined with Eqs.~(\ref{beta-via-alpha})
and (\ref{b-for-nonzero_ed-II}), yields
\begin{eqnarray}
b^{\dagger}(t)\!&=&\! h_1(t) +
                \lambda
                \sum_{k>0} \xi_k
                     \bigg[
                            g(\epsilon_k - i\eta)
                             (\epsilon_k + \omega_0)
                              e^{i \epsilon_k t}
                              \alpha^{\dagger}_k
\nonumber \\
               && \;\;\;\;\;\;\;\;\;\;\;\;\;\; -
                            g(\epsilon_k + i\eta)
                             (\epsilon_k-\omega_0)
                              e^{-i \epsilon_k t}
                             \alpha_k
                     \bigg]
\end{eqnarray}
with
\begin{eqnarray}
h_1(t) &=& \lambda \tilde{\epsilon}_d
           \sum_{k > 0} \xi_k^2
                      |g(\epsilon_k + i\eta)|^2
                      (\epsilon_k^2 - \omega_0^2)
\nonumber \\
     && \;\;\;\;\;\;\;\;\;\;\;\, \times
                      \left [
                              \frac{\epsilon_k + \omega_0}
                                   {\epsilon_k + i\eta}
                                   e^{i \epsilon_k t}
                            - \frac{\epsilon_k - \omega_0}
                                   {\epsilon_k - i\eta}
                                   e^{-i \epsilon_k t}
                    \right ]
\nonumber \\
     &+&
       \frac{\epsilon_d}{\pi \Gamma}
       \frac{g}{\omega_0}
       \frac{1}{1 - 2 g^2/(\pi \omega_0 \Gamma)} .
\end{eqnarray}
Accordingly, the phonon displacement is equal to
$Q(t) = \sqrt{2} {\rm Re} \{ h_1(t) \}$, which follows
from the fact that $\alpha_k$ and $\alpha_k^{\dagger}$
average to zero with respect to the initial state.
Similarly from Eq.~(\ref{n_d-tilde-via-beta}) one
has that
\begin{equation}
\tilde{n}_d(t) = \frac{1}{2} +
a \sum_{k>0}
        \xi_k (\epsilon_k^2-\omega_0^2)
        \left[
               g(\epsilon_k\!-\!i\eta)
               e^{i \epsilon_k t}
               \beta^{\dagger}_k
               + {\rm H.c.}
        \right]
\end{equation}
which, when combined with Eqs.~(\ref{beta-via-alpha})
and (\ref{n_d-nonzero-ed}), yields
\begin{equation}
\tilde{n}_d(t) = h_2(t) +
a \sum_{k>0}
        \xi_k (\epsilon_k^2-\omega_0^2)
        \left[
               g(\epsilon_k\!-\!i\eta)
               e^{i \epsilon_k t}
               \alpha^{\dagger}_k
               + {\rm H.c.}
        \right]
\end{equation}
with
\begin{eqnarray}
h_2(t) &=& \frac{1}{2} -
           \frac{\epsilon_d}{\pi \Gamma}
           \frac{1}{1 - 2 g^2/(\pi \omega_0 \Gamma)}
\nonumber \\
       &+&
           \tilde{\epsilon}_d a
           \sum_{k > 0} \xi_k^2
                      |g(\epsilon_k + i\eta)|^2
                      (\epsilon_k^2 - \omega_0^2)^2
\nonumber \\
     && \;\;\;\;\;\;\;\;\;\;\;\, \times
                      \left [
                              \frac{e^{i \epsilon_k t}}
                                   {\epsilon_k + i\eta}
                            + \frac{e^{-i \epsilon_k t}}
                                   {\epsilon_k - i\eta}
                    \right ] .
\label{h_2}
\end{eqnarray}
Hence the occupancy of the localized level is simply
given by $n_d(t) = h_2(t)$.

\begin{figure}[t]
\centerline{
\includegraphics[width=80mm]{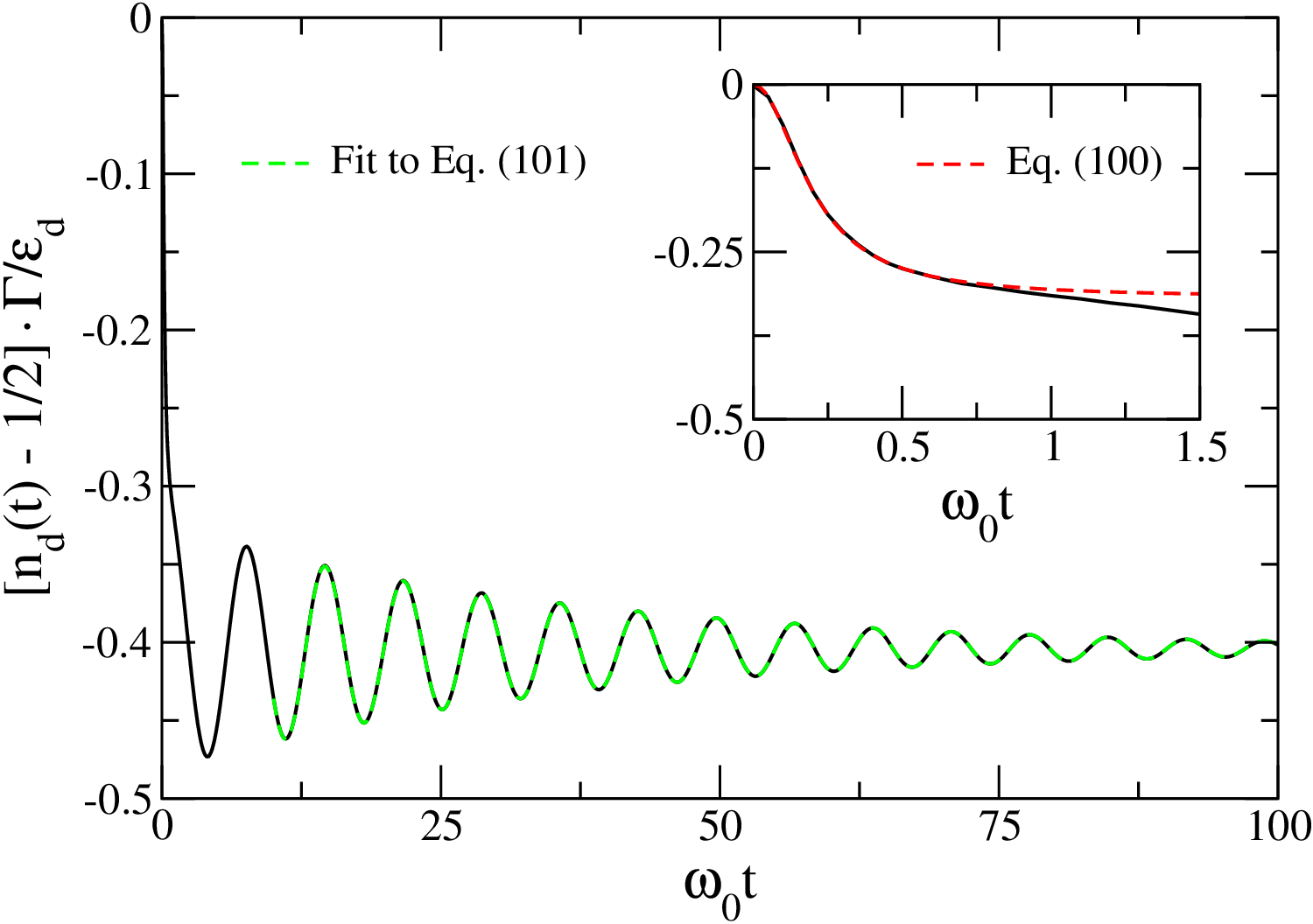}
}\vspace{0pt}
\caption{(Color online)
         Time evolution of the occupancy $n_d(t)$ of the
         localized electronic level, following an abrupt
         change in its energy from $\epsilon_d = 0$ to
         $\epsilon_d \neq 0$. Here $\omega_0/D_d = 0.2$
         and $g/\Gamma = 0.324$. Note that $n_d(t) - 1/2$
         depends linearly on $\epsilon_d$. The green
         dashed line shows a fit to Eq.~(\ref{n_d-fit})
         using the fitting range
         $10 \leq \omega_0 t \leq 100$.
         Inset: A zoom in on the short-time behavior.
         The red dashed curve shows the analytical
         form of Eq.~(\ref{n_d-short-t}).}
\label{Fig:n_d-Q3}
\end{figure}

The occupancy $n_d(t)$ of the localized electronic
level is depicted in Fig.~\ref{Fig:n_d-Q3}. Several
points are noteworthy. First, $n_d(t) - 1/2$ depends
linearly on $\epsilon_d$ in our solution. This
property stems from the fact that $\epsilon_d$
couples linearly to the bosonic degrees of
freedom, in accord with the assumption that
$|\epsilon_d| \ll \Gamma$. Indeed, as $|\epsilon_d|$
is increased the mapping onto the bosonic Hamiltonian
of Eq.~(\ref{H_bosonic}) gradually breaks down,
generating higher order corrections in $\epsilon_d$.

Second, the dynamics of $n_d(t)$ is composed of
two distinct segments: fast dynamics on the scale
of $1/D_d \sim 1/\Gamma$, where most of the charge
redistribution takes place, followed by an extended
region of damped oscillations. The short-time dynamics
originates from the high-energy end of the summation
over $k$ in Eq.~(\ref{h_2}), and is given by
\begin{equation}
n_d(t) = \frac{1}{2}
         - \frac{\epsilon_d}{\pi \Gamma}
           \frac{(t D_d)^2}{1 + (t D_d)^2}
\label{n_d-short-t}
\end{equation}
(see inset of Fig.~\ref{Fig:n_d-Q3}). This simple
analytical form stems from the exponential
high-energy cutoff imposed by $\xi_k^2$ [see
Eq.~(\ref{xi_q})]. While the functional form of
$n_d(t)$ may differ from Eq.~(\ref{n_d-short-t})
for other cutoff schemes, the relevant time scale
$t \sim 1/\Gamma$ and the characteristic change in
occupancy $\delta n_d \sim \epsilon_d/\pi \Gamma$
experienced within this time segment should be
generic. As for the damped oscillations, these
are expected to take the functional form
\begin{equation}
n_d(t) = A \sin(\Omega t + \phi) e^{-t/\tau_{0}} + C
\label{n_d-fit}
\end{equation}
with $\Omega$ and $\tau_0$ equal to $\omega$ and
$\tau$. We confirm this form in
Fig.~\ref{Fig:n_d-Q3}, where the fitted values
of $\Omega$ and $\tau_0$ agree to within 0.1\%
with those extracted from Fig.~\ref{Fig:Occ-Q1-n=0}
by fitting $n_b(t)$ to
Eq.~(\ref{n_b-fit}).~\cite{comment-on-fits}

\begin{figure}[t]
\centerline{
\includegraphics[width=75mm]{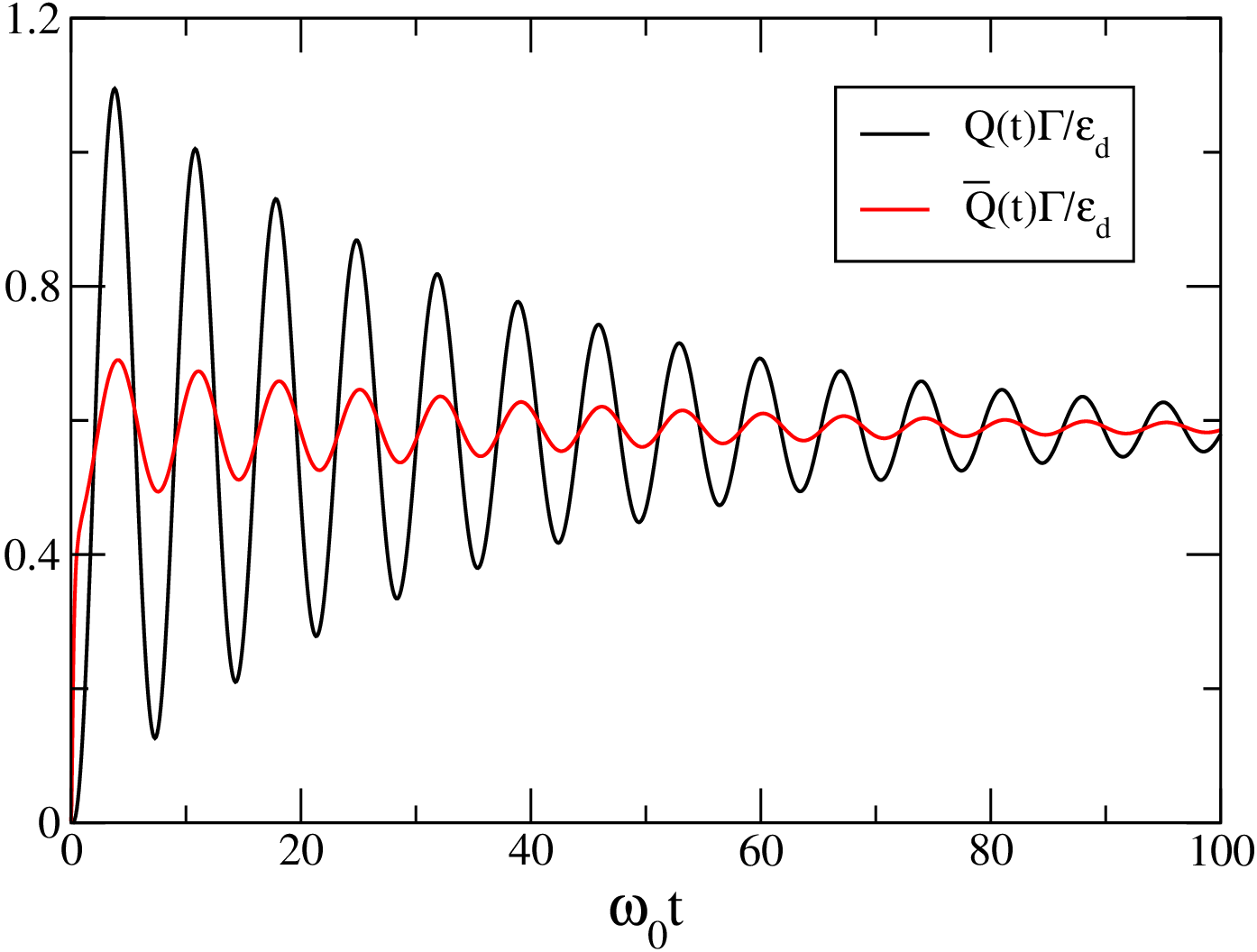}
}\vspace{0pt}
\caption{(Color online)
         Time evolution of the displacement $Q(t)$,
         following an abrupt change in the level
         energy from $\epsilon_d = 0$ to
         $\epsilon_d \neq 0$. All model parameters are
         the same as in Fig.~\ref{Fig:n_d-Q3}. For
         comparison, the red curve plots $\bar{Q}(t)$
         of Eq.~(\ref{bar-Q}).}
\label{Fig:Q-vs-t-Q3}
\end{figure}

Lastly, from the asymptotic long-time behavior of
$n_b(t)$ one can deduce the dimensionless parameter
controlling the perturbative expansion in $g$ in
thermal equilibrium. For $g = 0$ and arbitrary
$\epsilon_d$, the exact equilibrium occupancy of
the electronic level is given by the standard
expression
\begin{equation}
n_d = \frac{1}{2}
    - \frac{1}{\pi}
      \arctan \left (
                      \frac{\epsilon_d}{\Gamma}
              \right ) ,
\end{equation}
which reduces to $n_d = 1/2 - \epsilon_d/\pi \Gamma$
for $|\epsilon_d| \ll \Gamma$. This latter
result is accurately reproduced by our treatment
upon setting $g$ equal to zero. For nonzero $g$
we find that
\begin{equation}
n_d = \frac{1}{2} -
      \frac{\epsilon_d}{\pi \Gamma}
      \frac{1}{1 - 2 g^2/(\pi \omega_0 \Gamma)} ,
\end{equation}
revealing that the true expansion parameter is
$g^2/\omega_0 \Gamma$, i.e., the ratio of the
polaronic shift $g^2/\omega_0$ to the hybridization
width $\Gamma$. The effect of the electron-phonon
coupling $g$ is to increase the deviation from
half filling for a given value of $\epsilon_d$,
signaling a narrowing of the electronic resonance
according to
\begin{equation}
\Gamma \to \Gamma_{\rm eff} =
       \Gamma - \frac{2 g^2}{\pi \omega_0} .
\end{equation}

The time evolution of the phonon displacement
is plotted in turn in Fig.~\ref{Fig:Q-vs-t-Q3}.
Similar to the level occupancy, $Q(t)$ depends
linearly on $\epsilon_d$ and undergoes damped
oscillations with the relaxation time $\tau$ and
frequency $\omega$. It lacks, however, the fast
dynamics that the level occupancy experiences
on the time scale of $1/\Gamma$. In equilibrium
$n_d$ and $Q$ are related through
\begin{equation}
Q = -\sqrt{2} \frac{g}{\omega_0}
              \left [
                      n_d - \frac{1}{2}
              \right ] ,
\label{Q-via-n_d}
\end{equation}
which is an exact result applicable to
arbitrary $\epsilon_d$, $\omega_0$, and
$g$.~\cite{comment-on-identity} It is thus
natural to ask whether this general relation
extends to nonequilibrium dynamics. To this end,
in Fig.~\ref{Fig:Q-vs-t-Q3} we have plotted
\begin{equation}
\bar{Q}(t) = -\sqrt{2}
              \frac{g}{\omega_0}
              \left [
                      n_d(t) - \frac{1}{2}
              \right ]
\label{bar-Q}
\end{equation}
alongside $Q(t)$. Although nearly in phase, the
two quantities are characterized by vastly different
amplitudes of oscillations, marking the breakdown of
Eq.~(\ref{Q-via-n_d}) under nonequilibrium dynamics.
The latter relation is restored only asymptotically
as the system thermalizes.

\section{Driven dynamics}
\label{sec:driven-dynamics}

Up until now we considered the response to a single
quantum quench. Quantum control of nanodevices often
requires the usage of driven dynamics, where periodic
forcing is applied to the system. Such drives are a
theoretical challenge to describe since the system not
only remains permanently remote from thermal equilibrium,
but it never even reaches steady state. Remarkably,
we are able to extend our exact solution to a
rather broad class of driven dynamics where the
forcing couples linearly to the bosonic degrees
of freedom. As we discuss below, this class of
drives includes at least two physically relevant
scenarios where periodic forcing is applied either
to the localized phonon or to the electronic level.
Accordingly, we begin our derivation with a general
discussion of this class of drives before turning
to the two concrete examples of interest.

\subsection{Drives that couple linearly to bosons}

The general setting we consider consists of a system
that resides at time $t < 0$ in thermal equilibrium,
when a time-dependent drive is suddenly applied to
it. In formal terms, the Hamiltonian of the system
is changed abruptly at time $t = 0$ from the
Hamiltonian ${\cal H}$ of Eq.~(\ref{H_bosonic}) to
${\cal H}'(t) = {\cal H} + {\cal H}_{\rm drive}(t)$
with
\begin{equation}
{\cal H}_{\rm drive}(t) =
         \sum_{k > 0}
         \left [
                 M_k(t) \alpha^{\dagger}_k
                 + M_k^{\ast}(t) \alpha_k
         \right ] .
\label{H_drive}
\end{equation}
Here we have assumed that the drive couples linearly
to the eigenmodes of ${\cal H}$, exploiting the
fact that any linear combination of the original
bosonic degrees of freedom can be expanded in terms
of the scattering-state operators $\alpha_k$ and
$\alpha^{\dagger}_k$. The coefficients $M_k(t)$
depend on the exact scenario under consideration
and will typically have the separable form
$M_k(t) = A(t) m_k$. Nevertheless, we shall regard
them for the time being as general coefficients
without making any further assumption about their form.
The initial value of the energy level $\epsilon_d$
will be taken for simplicity to be zero, though
the extension to nonzero $\epsilon_d$ is quite
straightforward.

A convenient way to incorporate the time-dependent
drive is via the Heisenberg equation of motion for
the scattering-state operators, which takes the
form
\begin{equation}
\dot{\alpha}_k(t) = -i\epsilon_k \alpha_k(t)
                    - i M_k(t) ,
\label{EOM-w-drive}
\end{equation}
subject to the initial condition
$\alpha_k(t = 0) = \alpha_k$. Here the first term
on the right-hand side of Eq.~(\ref{EOM-w-drive})
is due to ${\cal H}$ and the second term is due to
${\cal H}_{\rm drive}$. Equation~(\ref{EOM-w-drive})
has the formal solution
\begin{equation}
\alpha_k(t) = \alpha_k e^{-i\epsilon_k t}
            - i \int_{0}^{t}\!
                     e^{i\epsilon_k (t' - t)}
                     M_k(t') d t' ,
\label{alpha_k-drive}
\end{equation}
from which the time evolution of all physical operators
of interest can be deduced. For example, combining
Eq.~(\ref{b_via_sc_state}) with Eq.~(\ref{alpha_k-drive})
and its Hermitian conjugate one obtains
\begin{equation}
b^{\dagger}(t) = b_0^{\dagger}(t) + i \lambda {\cal B}(t) ,
\end{equation}
where $b_0^{\dagger}(t)$ is the time-evolved
operator in the absence of a drive [see
Eq.~(\ref{b_time_dep})] and ${\cal B}(t)$ is
a time-dependent shift given by
\begin{eqnarray}
{\cal B}(t)\!&=&\!\sum_{k>0} \xi_k
                     \bigg[
                            g(\epsilon_k\!-\!i\eta)
                             (\epsilon_k\!+\!\omega_0)
                             \int_{0}^{t}\!\!
                             M_k^{\ast}(t')
                             e^{-i\epsilon_k (t' - t)}
                             d t'
\nonumber \\
            && +
                            g(\epsilon_k\!+\!i\eta)
                             (\epsilon_k\!-\!\omega_0)
                             \int_{0}^{t}\!\!
                             M_k(t')
                             e^{i\epsilon_k (t' - t)}
                             d t'
                     \bigg] .
\end{eqnarray}
Accordingly, the phonon displacement takes the form
\begin{equation}
Q(t) = -\sqrt{2} \lambda
        {\rm Im} \{ {\cal B}(t) \} ,
\label{Q-general-drive}
\end{equation}
while its occupancy reads
\begin{equation}
n_b(t) = n_b^{(0)} + \lambda^2 |{\cal B}(t)|^2 .
\label{n_b-general-drive}
\end{equation}
Here $n_b^{(0)}$ denotes the equilibrium phononic
occupancy in the absence of a drive, given by
Eq.~(\ref{eq_occ_infty}) for $T = 0$.
Note that in deriving Eqs.~(\ref{Q-general-drive})
and (\ref{n_b-general-drive}) we have made use of
the fact that $b(t) + b^{\dagger}(t)$ and
$b^{\dagger}(t) b(t)$ are averaged with respect to
the equilibrium density operator corresponding to
${\cal H}$, which is diagonal in the occupation
numbers $\alpha_k^{\dagger}\alpha_k$. As a result
$\langle \alpha_k \rangle$ and
$\langle \alpha^{\dagger}_k \rangle$ identically
vanish. A similar calculation for the time-dependent
occupancy $\delta n_d(t) = n_d(t) - 1/2$ of the
localized level yields
\begin{eqnarray}
\delta n_d \! &=& \!\! a \sum_{k > 0}
             \xi_k (\epsilon_k^2\!-\!\omega_0^2)
             \bigl [
                     -i g(\epsilon_k\!+\!i\eta)
                     \int_{0}^{t}\!\!
                                 M_k(t')
                                 e^{i \epsilon_k (t' - t)}
                                 d t'
\nonumber \\
               &&\!\!
                + \; i g(\epsilon_k - i\eta)
                     \int_{0}^{t}\!
                                 M_k^{\ast}(t')
                                 e^{-i \epsilon_k (t' - t)}
                                 d t'
             \bigr ] .
\label{n_d-general-drive}
\end{eqnarray}

Equations~(\ref{n_b-general-drive}) and
(\ref{n_d-general-drive}) combined provide us with
a formal solution for the occupancies of the local
phonon and the electronic level for a general drive.
Furthermore, these expressions apply to arbitrary
temperature $T$, provided $n^{(0)}_b$ is taken to be
the equilibrium phononic occupancy at that temperature.
Below we utilize these expressions to analyze two cases
of practical interest, where ac forcing is applied
either to the local phonon or to the localized level.

\subsection{ac forcing of the local phonon}

In the first scenario to be analyzed, ac forcing is
applied at time $t > 0$ to the local phonon, as
described by the Hamiltonian term
\begin{equation}
{\cal H}_{\rm drive}(t) = \Delta \sin(\Omega t)
                          (b^{\dagger} + b) .
\label{H-drive-ac-p}
\end{equation}
Here $\Delta$, which has dimensions of energy, is
the amplitude of the drive and $\Omega$ is the
driving frequency (not to be confused with the
fitting parameter previously used for analyzing
the damped oscillations). Such forcing can be
applied, e.g., to polar molecules using an ac
electric field.

Using the expansion of $b^{\dagger}$ in
terms of the scattering-state operators given in
Eq.~(\ref{b_via_sc_state}), the Hamiltonian term
of Eq.~(\ref{H-drive-ac-p}) can be recast in the
form of Eq.~(\ref{H_drive}) with the coefficients
\begin{equation}
M_k(t) = 2 \omega_0 \lambda \Delta \sin(\Omega t)
         \xi_k g(\epsilon_k - i\eta) ,
\end{equation}
such that
\begin{equation}
\int_{0}^{t}\!
         M_k^{\ast}(t') e^{-i \epsilon_k (t' - t)} d t' =
         \omega_0 \lambda \Delta
         \xi_k g(\epsilon_k + i\eta)
         \zeta_k(t)
\label{int-of-M^ast}
\end{equation}
with
\begin{equation}
\zeta_k(t) =
             \frac{e^{i\Omega t} - e^{i\epsilon_k t}}
                  {\epsilon_k - \Omega}
             - \frac{e^{-i\Omega t} - e^{i\epsilon_k t}}
                    {\epsilon_k + \Omega} .
\label{zeta_without_eta}
\end{equation}
For computational convenience it is useful to add
an infinitesimal imaginary part $i\eta$ to the
denominators in Eq.~(\ref{zeta_without_eta}), thereby
rewriting $\zeta_k(t)$ as~\cite{comment-on-eta}
\begin{equation}
\zeta_k(t) =
             \frac{e^{i\Omega t} - e^{i\epsilon_k t}}
                  {\epsilon_k - \Omega + i\eta}
             - \frac{e^{-i\Omega t} - e^{i\epsilon_k t}}
                    {\epsilon_k + \Omega + i\eta} .
\label{zeta_with_eta}
\end{equation}
A somewhat lengthy calculation then gives
\begin{equation}
\lambda {\cal B}(t) = \frac{\Delta}{2}
                      \left [
                              F(-\Omega - i\eta, t)
                              - F(\Omega - i\eta, t)
                      \right ] ,
\end{equation}
where $F(z, t)$ is the same function defined in
Eq.~(\ref{eq_f_defined}). The phonon displacement
and occupancy therefore take the rather compact
forms
\begin{equation}
Q(t) = -\frac{\Delta}{\sqrt{2}}
        {\rm Im} \{
                    F(-\Omega - i\eta, t)
                    - F(\Omega - i\eta, t)
                 \}
\label{Q-ac-drive}
\end{equation}
and
\begin{equation}
n_b(t) = n_b^{(0)} + \frac{\Delta^2}{4}
         | F(-\Omega - i\eta, t)
           - F(\Omega - i\eta, t) |^2 .
\label{n_b-ac-drive}
\end{equation}

\begin{figure}[t]
\centerline{
\includegraphics[width=80mm]{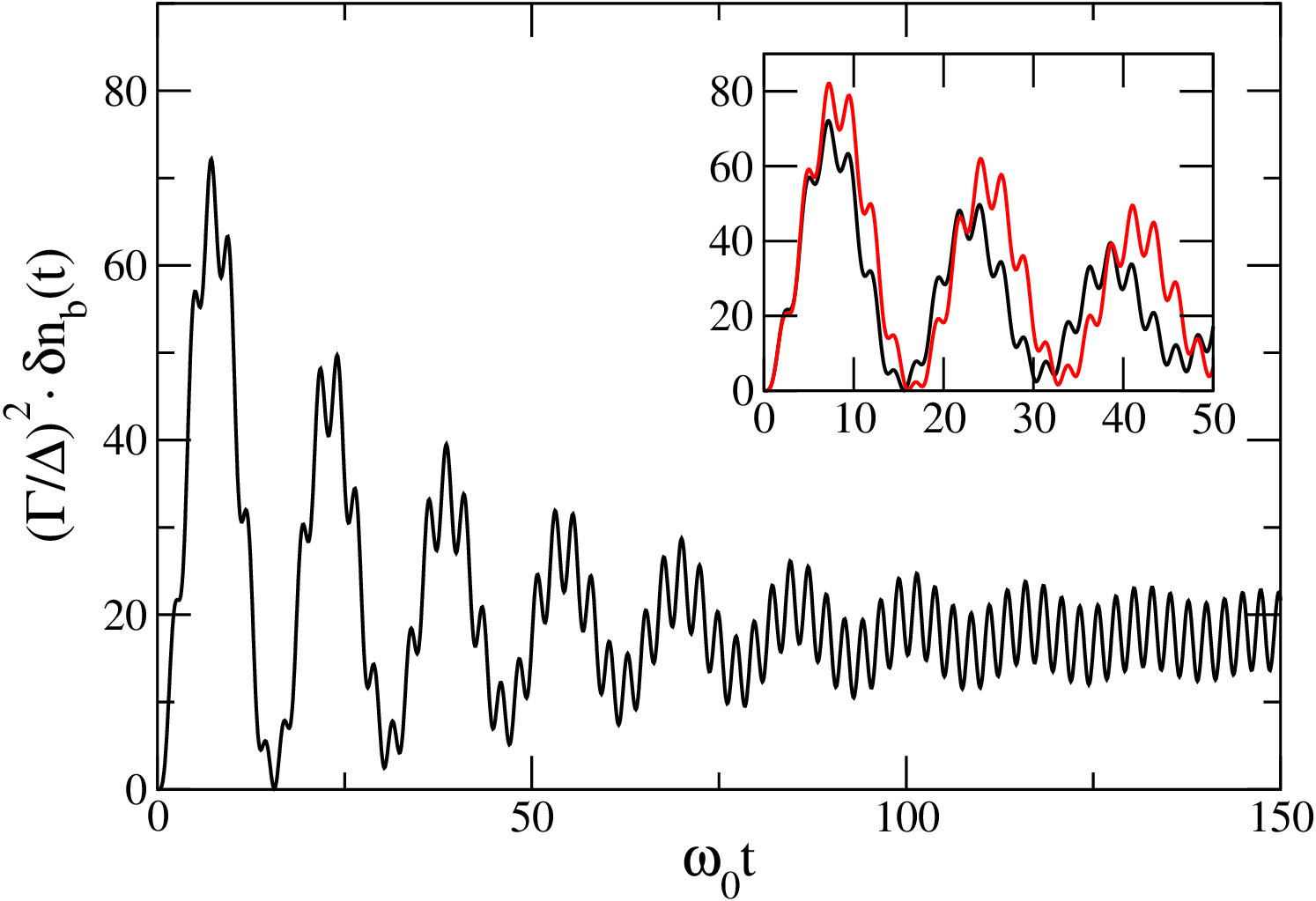}
}\vspace{0pt}
\caption{(Color online)
         Time evolution of $\delta n_b(t) = n_b(t)
         - n_b^{(0)}$ in response to ac forcing of the
         phonon according to ${\cal H}_{\rm drive}(t)$
         of Eq.~(\ref{H-drive-ac-p}). Here
         $\omega_0/D_d = 0.2$, $g/\Gamma = 0.28$,
         and $\Omega = 1.3 \omega_0$. Initially there
         is a rich structure involving the interference
         of four distinct frequencies. As transients
         decay (on a time scale of $\tau$),
         $\delta n_b(t)$ gradually reduces to a
         single harmonic with a frequency
         of oscillations equal to $2\Omega$.
         Inset: Zoom in on the earlier time segment
         $\omega_0 t < 50$, including a comparison
         to the stronger coupling strength
         $g/\Gamma = 0.324$ (red curve).}
\label{Fig:n_b-ac-1-off-res}
\end{figure}

The general structure of $Q(t)$ and $n_b(t)$ could
be understood from properties of the function
$F(z, t)$. Since $F(z, 0)$ identically vanishes for
arbitrary $z$, the phononic occupancy and displacement
properly reduce at time $t = 0$ to their thermal
equilibrium values, as they physically should.
As soon as $t > 0$, the two components of
$F(\pm \Omega - i\eta)$ behave markedly differently.
The first term in Eq.~(\ref{eq_f_defined})
oscillates indefinitely with frequency $\Omega$,
whereas the term involving the sum over $k$ undergoes
damped oscillations with the relaxation time $\tau$
and frequency $\omega$ of Eqs.~(\ref{tau}) and
(\ref{omega}), respectively. Thus, there is a clear
distinction between the roles of the two terms:
while the first term in Eq.~(\ref{eq_f_defined})
survives at long times and is responsible for the
long-time behavior, the second term contains all
transients that decay in time. This leads to the
following characterization of $Q(t)$ and $n_b(t)$.
At short times, $t < \tau$, the phonon displacement
comprises of two components oscillating at
frequencies $\Omega$ and $\omega$, while $n_b(t)$
contains four distinct oscillatory terms with the
frequencies $2\Omega$, $\Omega \pm \omega$, and
$2\omega$. At long times, $\tau \ll t$, the phonon
displacement oscillates with frequency $\Omega$
about zero, while $n_b(t)$ oscillates with
frequency $2\Omega$ about a new time-averaged
value $\bar{n}_b$. Explicitly, $Q(t)$ and
$\delta n_b(t) = n_b(t) - n_b^{(0)}$ reduce at
long times to
\begin{equation}
Q(t) = \sqrt{2} \Delta \omega_0
                |g(\Omega + i\eta)|
                \sin (\Omega t + \phi)
\end{equation}
and
\begin{equation}
\delta n_b(t)\!=\!
       \frac{\Delta^2}{2} |g(\Omega + i\eta)|^2
             \bigl [
                     \omega_0^2 + \Omega^2
                     + \left (
                               \Omega^2\!-\!\omega_0^2
                       \right )
                       \cos( 2\Omega t + 2 \phi)
             \bigr ] ,
\label{n_b-ac-1-long-t}
\end{equation}
with $\phi = \arg \{ g(\Omega - i\eta) \}$.

On physical grounds one expects the response to an ac
drive to reduce at long times to a periodic function
of time, containing all harmonics of the driving
frequency $\Omega$. Surprisingly, the oscillatory
parts of $Q(t)$ and $\delta n_b(t)$ consist in this
limit of just a single harmonic each.
While $Q(t)$ tracks the driving field with a phase
difference of $\phi$, $\delta n_b(t)$ oscillates
with the doubled frequency $2\Omega$, lacking any
signal at the principal harmonic $\Omega$. Such
behavior is quite atypical, as is the absence of
higher harmonics. We expect both features to
qualitatively change as the electron-phonon coupling
is increased beyond the validity of our solution.
We further note that the doubling of frequency
in $\delta n_b(t)$ is reminiscent of a similar
doubling of frequency in the damped oscillations
that $n_b(t)$ undergoes in response to a quantum
quench (see, e.g., Figs.~\ref{Fig:Occ-Q1-n=0} and
\ref{Fig:Q-vs-t-Q1} and their accompanying texts).

\begin{figure}[t]
\centerline{
\includegraphics[width=80mm]{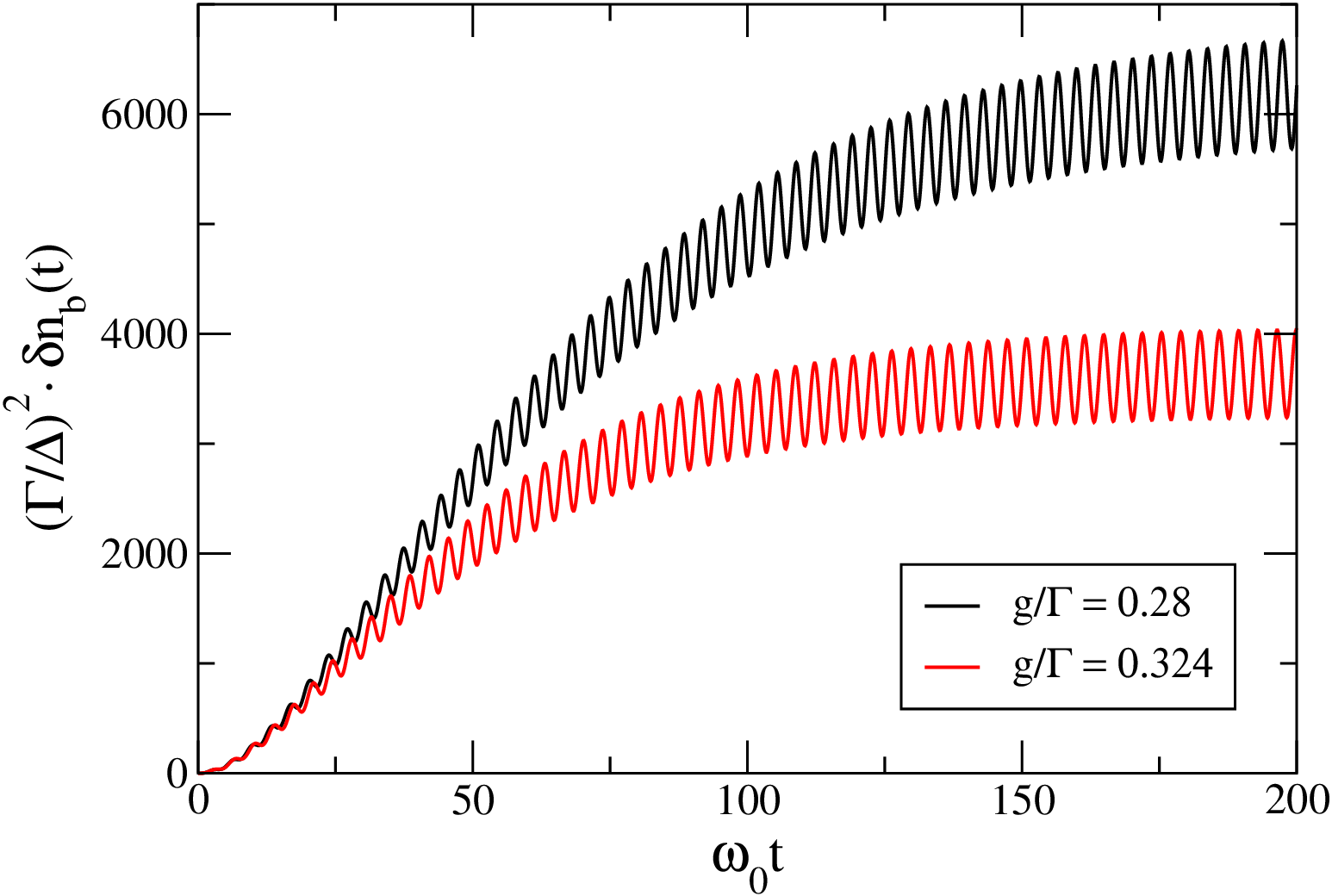}
}\vspace{0pt}
\caption{(Color online)
         Same as Fig.~\ref{Fig:n_b-ac-1-off-res}, with
         $\Omega$ tuned to the resonance frequency:
         $\omega = 0.896\omega_0$ for
         $g/\Gamma = 0.324$ and $\omega
         = 0.923\omega_0$ for $g/\Gamma = 0.28$.
         All other model parameters are the same as
         in Fig.~\ref{Fig:n_b-ac-1-off-res}. Note
         the vastly different vertical scale as
         compared to that used in
         Fig.~\ref{Fig:n_b-ac-1-off-res}.}
\label{Fig:n_b-ac-1-res}
\end{figure}

Focusing on $\delta n_b(t)$, its amplitude depends
in a simple quadratic manner on $\Delta$. Other
than setting the overall amplitude, $\Delta$ has
no additional effect on $\delta n_b(t)$. By
contrast, the shape of $\delta n_b(t)$ is quite
sensitive to the driving frequency $\Omega$, as
demonstrated in Figs.~\ref{Fig:n_b-ac-1-off-res}
and \ref{Fig:n_b-ac-1-res}. When $\Omega$ is tuned
off-resonance with the intrinsic frequency $\omega$
of the system, see Fig.~\ref{Fig:n_b-ac-1-off-res},
the transient behavior shows a rather rich structure
that stems from the interference of the four underlying
frequencies $2\Omega$, $\Omega \pm \omega$, and
$2\omega$. Only after all transients have decayed
on a time scale of $\tau$ does $\delta n_b(t)$
approach its asymptotic long-time form of a single
harmonic with the doubled frequency $2\Omega$.

\begin{figure}[t]
\centerline{
\includegraphics[width=80mm]{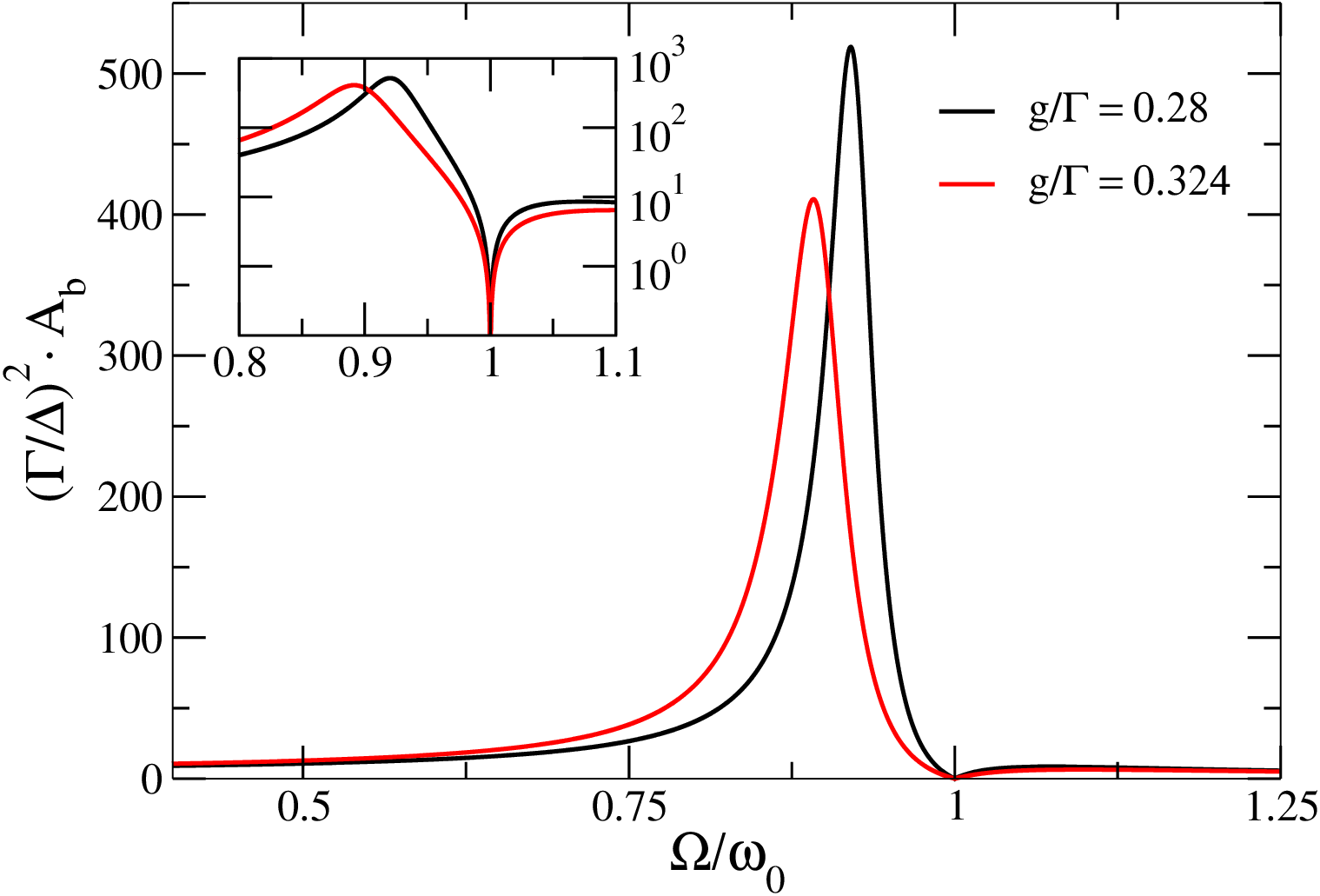}
}\vspace{0pt}
\caption{(Color online)
         The asymptotic long-time amplitude of
         oscillations $A_b$ vs the driving frequency
         $\Omega$, for $\omega_0/D_d = 0.2$ and two
         different strengths of the electron-phonon
         coupling $g$.
         Inset: A zoom in on the vicinity of the
         resonance peak. A logarithmic scale is used
         for the $y$ ordinate so as to emphasize the
         vanishing of $A_b$ for $\Omega = \omega_0$.}
\label{Fig:Amp-ac-1}
\end{figure}

A rather different picture is recovered when
$\Omega$ is tuned to the resonance frequency
$\omega$, see Fig.~\ref{Fig:n_b-ac-1-res}. Here
both the transient and long-time behaviors are
governed by the same single frequency $2\Omega =
2\omega$, resulting in much smoother curves.
Quite striking is the substantial increase of
the amplitude of oscillations upon approaching
the resonance frequency. Indeed, the amplitude of
the long-time oscillations is roughly two orders
of magnitude larger in Fig.~\ref{Fig:n_b-ac-1-res}
as compared to Fig.~\ref{Fig:n_b-ac-1-off-res}, which
is readily understood from Eq.~(\ref{n_b-ac-1-long-t}).
Since the amplitude of oscillations is given at long
times by
\begin{equation}
A_b = \frac{\Delta^2}{2} |g(\Omega + i\eta)|^2
      |\omega_0^2 - \Omega^2| ,
\label{A_b}
\end{equation}
it displays a sharp resonance for $\Omega \approx \omega$
where $|g(\Omega + i\eta)|$ is sharply
peaked.~\cite{comment-on-A_b} The amplitude of
oscillations at resonance can be crudely estimated as
\begin{equation}
A^{\rm res}_b \sim \frac{(\Delta \tau)^2}{8 \omega^2}
                   |\omega_0^2 - \omega^2| ,
\end{equation}
with $\omega$ and $\tau$ approximately given by
Eqs.~(\ref{omega}) and (\ref{tau}), respectively.
Another interesting observation is the vanishing
of $A_b$ for $\Omega = \omega_0$. A plot of the
amplitude $A_b$ of the long-time oscillations
as a function of $\Omega$ is depicted in
Fig.~\ref{Fig:Amp-ac-1}.

\subsection{ac forcing of local electronic level}

In the second scenario that we analyze, ac
forcing is applied at time $t > 0$ to the
localized electronic level, as described by
the Hamiltonian term
\begin{equation}
{\cal H}_{\rm drive}(t) =
         \Delta \sin(\Omega t)
         \left (
                 \hat{n}_d - \frac{1}{2}
         \right ) .
\label{H-drive-ac-e}
\end{equation}
Here, as before, $\Delta$ denotes the amplitude of
the drive and $\Omega$ is the forcing frequency.
Experimentally such a drive can be realized by
applying microwave voltage to a near-by plunger
gate, similar to the setups used by
Elzerman {\em et al}.~\cite{Elzerman-ac-00}
and by Kogan {\em et al}.~\cite{Kogan-ac-04} in
their respective studies of the ac Kondo effect in
semiconductor quantum dots.

Using the mode expansion of Eq.~(\ref{n_d-via-alpha})
one can again recast the Hamiltonian term of
Eq.~(\ref{H-drive-ac-e}) in the form of
Eq.~(\ref{H_drive}), this time with the
coefficients
\begin{equation}
M_k(t) = \Delta a \sin(\Omega t)
         \xi_k g(\epsilon_k - i\eta)
         (\epsilon_k^2 - \omega_0^2) .
\end{equation}
Repeating the same sequence of steps detailed in
Eqs.~(\ref{int-of-M^ast})--(\ref{zeta_with_eta})
and plugging the resulting expressions into
Eq.~(\ref{n_d-general-drive}), one obtains
after a rather lengthy calculation
\begin{equation}
\delta n_d (t) \equiv n_d (t) - \frac{1}{2}
               = \frac{\Delta}{g^2}
                 {\rm Im} \{
                             \tilde{F}(\Omega -i\eta,t)
                          \}
\end{equation}
with
\begin{eqnarray}
\tilde{F}(z, t) &=&
      (z^2 - \omega_0^2) g(z) \Sigma(z) e^{i z t}
      + \lambda^2 \sum_{k > 0}
                       \xi_k^2 |g(\epsilon_k + i\eta)|^2
\nonumber \\
                && \times
                   \left(
                          z^2 - \omega_0^2
                   \right)^2
                   \left(
                          \frac{e^{i \epsilon_k t}}
                               {\epsilon_k - z}
                          + \frac{e^{-i \epsilon_k t}}
                                 {\epsilon_k + z}
                   \right) .
\end{eqnarray}

\begin{figure}[t]
\centerline{
\includegraphics[width=80mm]{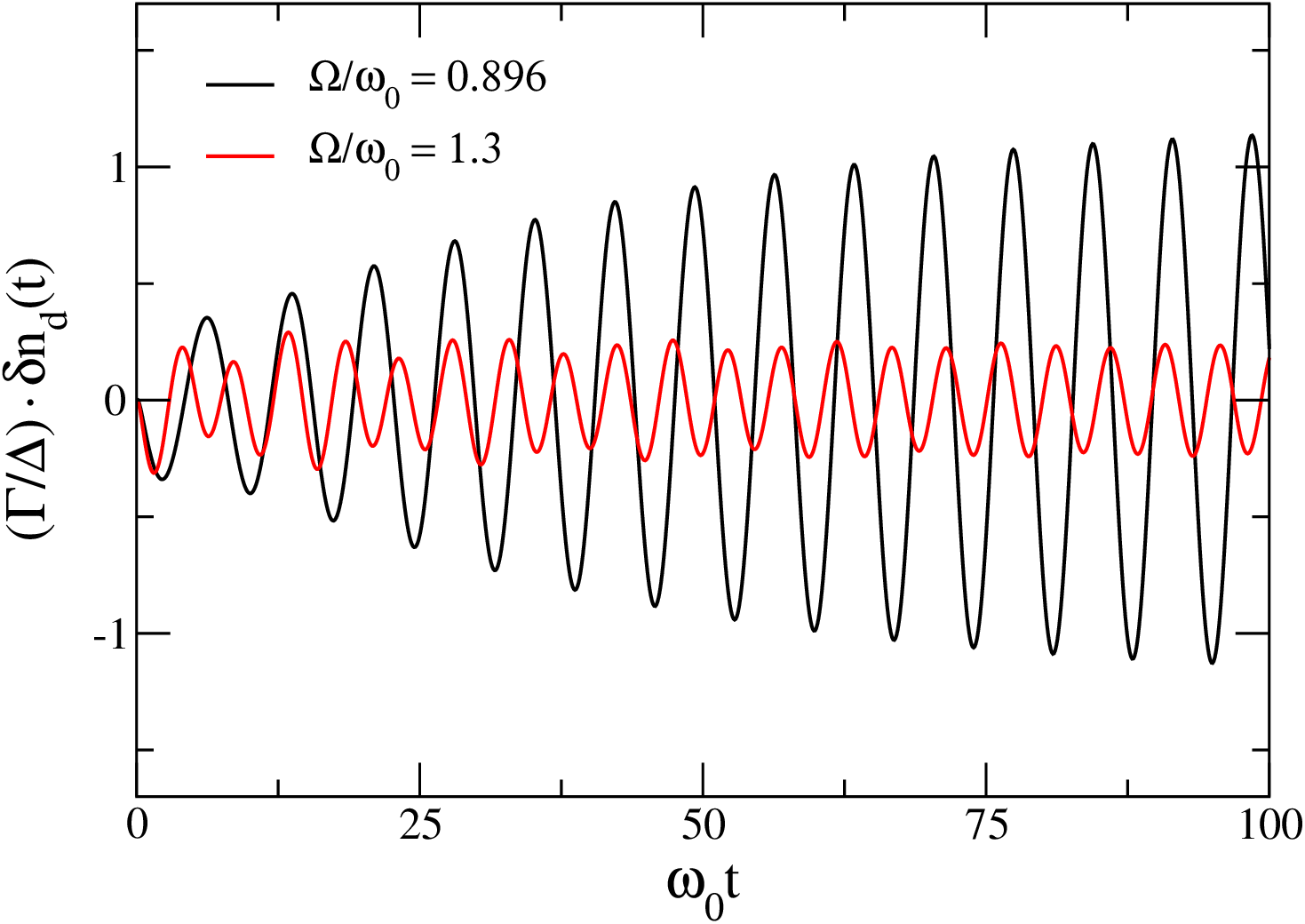}
}\vspace{0pt}
\caption{(Color online)
         Time evolution of $\delta n_d(t)$ in response
         to ac forcing of the local electronic level
         according to ${\cal H}_{\rm drive}(t)$
         of Eq.~(\ref{H-drive-ac-e}). Here
         $\omega_0/D_d = 0.2$ and $g/\Gamma = 0.324$.
         Two driving frequencies are shown, one
         ($\Omega = 1.3 \omega_0$) off resonance and
         the other ($\Omega = 0.896\omega_0$) on
         resonance with the internal frequency
         $\omega$.}
\label{Fig:n_d-ac-2}
\end{figure}

\begin{figure}[t]
\centerline{
\includegraphics[width=80mm]{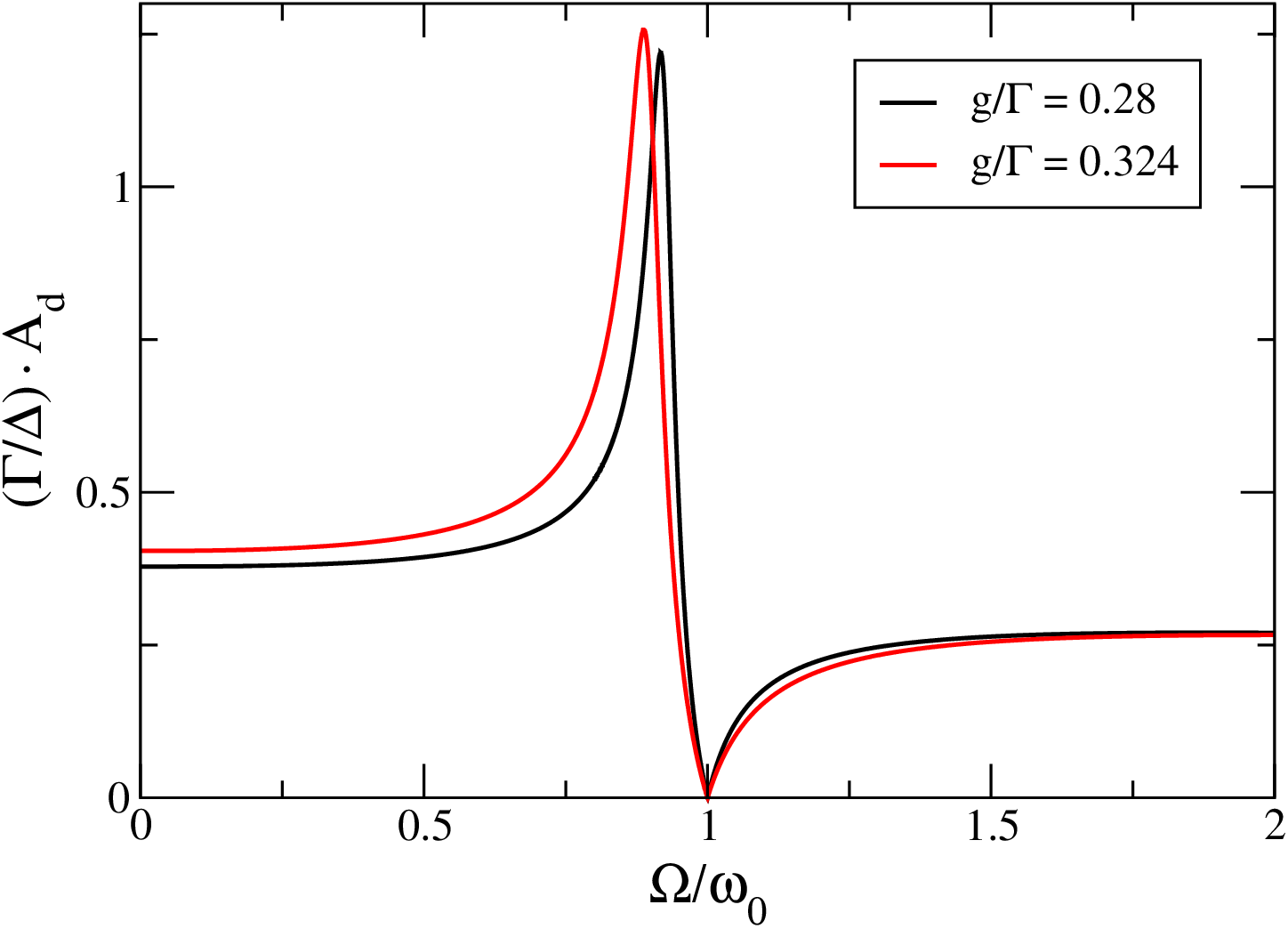}
}\vspace{0pt}
\caption{(Color online)
         The asymptotic long-time amplitude of
         oscillations $A_d$ vs the driving frequency
         $\Omega$, for $\omega_0/D_d = 0.2$ and two
         different strengths of the electron-phonon
         coupling $g$.}
\label{Fig:Amp-ac-2}
\end{figure}

The function $\tilde{F}(z, t)$ has similar
properties to those of $F(z, t)$. At $t = 0$ it
vanishes identically for any value of $z$, and
is composed of two distinct components for
$t > 0$: one that oscillates indefinitely with
frequency $\Omega$, and another that oscillates
with frequency $\omega$ and decays with the
relaxation time $\tau$ for any $z$ in the
lower half plane. Since $\delta n_d (t)$
is proportional to the imaginary part of
$\tilde{F}(\Omega - i\eta, t)$ [as opposed to
$\delta n_b(t)$ that depends quadratically on
$F(\pm\Omega - i\eta, t)$] it comprises at short
times $t < \tau$ of two oscillatory terms, one
with frequency $\Omega$ and another with frequency
$\omega$. As $t$ exceeds $\tau$ the latter component
is progressively suppressed and $\delta n_d (t)$
gradually approaches its asymptotic long-time form
\begin{equation}
\delta n_d (t) = A_d \sin (\Omega t + \varphi) ,
\end{equation}
with
\begin{equation}
A_d = \frac{\Delta}{g^2}
           |(\Omega^2 - \omega_0^2) g(\Omega + i \eta)
            \Sigma(\Omega + i \eta)|
\end{equation}
and $\varphi = \arg \left \{ (\Omega^2 - \omega_0^2)
g(\Omega - i \eta) \Sigma(\Omega - i \eta) \right \}$.
As beforehand, the long-time oscillations develop a
resonance for $\Omega \approx \omega$, albeit
with a reduced amplitude as compared to $A_b$ of
Eq.~(\ref{A_b}). This reduction in amplitude stems
from a weaker linear dependence of $A_d$ on
$|g(\Omega + i\eta)|$. Similar to $A_b$, the
amplitude of oscillations is suppressed to zero
for $\Omega = \omega_0$, leaving no signal at
long times for this particular frequency. A
summary of our results is presented in
Figs.~\ref{Fig:n_d-ac-2} and \ref{Fig:Amp-ac-2}.

\section{Summary and conclusions}
\label{sec:conclusions}

In this paper we have presented an asymptotically
exact solution for the nonequilibrium dynamics of
a single-molecule transistor in response to various
quantum quenches and drives. Our solution, which is
based on a controlled mapping of the original
Hamiltonian of Eq.~(\ref{H_full}) onto a form quadratic
in bosonic operators,~\cite{Dora_Halbritter_2009}
is formally confined to weak electron-phonon
coupling and near-resonance conditions for the
electronic level:
$\Gamma \gg \max\{g, |\epsilon_d|, g^2/\omega_0 \}$.
While some aspects of this regime can be accessed
using ordinary perturbation theory in $g$, the ability
to sum all orders exactly allowed us to (i)
explicitly show how the system thermalizes following
a quantum quench, (ii) identify the different time
scales that govern the dynamics of the system, and
(iii) access the asymptotic long-time response to
a periodic drive.

Transient behaviors following a quantum quench
were found to involve two characteristic
scales~\cite{Dora_Halbritter_2009} --- an
intrinsic frequency $\omega$ and a relaxation
time $\tau$ approximately given by
Eqs.~(\ref{omega}) and (\ref{tau}), respectively.
Quite surprisingly, some observables, such
as the phonon displacement and the electronic
occupancy of the localized level, display damped
oscillations with frequency $\omega$ and the relaxation
time $\tau$, while other observables, such as the
phononic occupancy, oscillate with frequency $2\omega$
and decay with the reduced relaxation time $\tau/2$.
The distinction has to do with the bosonic
representation of the observable in question. If
the latter is expressed as a linear combination of
bosonic operators, the relevant frequency and decay
time are $\omega$ and $\tau$. If, on the other hand,
the the observable in question is quadratic in bosonic
operators, the relevant frequency and decay time are
$2\omega$ and $\tau/2$, respectively.

A special feature of our solution is the nature of
the long-time response of observables to ac drives,
whose oscillatory component reduces to just a
single harmonic with an amplitude that depends
in a simple power-law fashion on the forcing
amplitude $\Delta$. The absence of additional
harmonics in the long-time ac response is a
direct consequence of the mapping onto a free
bosonic Hamiltonian with a forcing field that couples
linearly to the bosonic modes. Physically this
implies that other harmonics, which are generally
expected to exist for the original Hamiltonian of
Eq.~(\ref{H_full}), are parametrically small for
$\Gamma \gg \max \{ g, g^2/\omega_0 \}$. As the
electron-phonon interaction is increased such
that $\max \{ g, g^2/\omega_0 \}$ approaches
$\Gamma$, additional harmonics are expected to
gain significance, that is provided the forcing
amplitude $\Delta$ is not too small. Concomitantly,
the amplitudes of the different harmonics should
gradually acquire a more elaborate dependence on
$\Delta$ beyond a simple power-law form. We
emphasize that this regime can no longer be
described by the bosonic Hamiltonian of
Eq.~(\ref{H_bosonic}).

It would be interesting to compare our results
with a numerical evaluation of the quench
dynamics using, e.g., the time-dependent
numerical renormalization group
(TD-NRG).~\cite{TD-NRG-PRL,TD-NRG-PRB}
Since $\omega$ and $\tau$ are independent of the
high-energy cutoff used in the electronic Hamiltonian
of Eqs.~(\ref{H_full}) and (\ref{H_0}), this
should facilitate a
direct comparison between the two approaches on
time scales exceeding $1/D_d \sim 1/\Gamma$. The
precise forms of $\omega$ and $\tau$, as well as
the short-time dynamics up to $t \sim 1/\Gamma$,
do depend on the cutoff scheme used for the
bosonized Hamiltonian of Eq.~(\ref{H_bosonic}).
Nevertheless, we expect our weak-coupling
expressions for $\omega$ and $\tau$ to apply
in their present forms, as these coincide with
low-order perturbation theory in $g$ when
applied directly to the electronic Hamiltonian
of Eq.~(\ref{H_full}).

The true power of the TD-NRG lies, however, in its
ability to treat arbitrary couplings strengths,
which should enable one to go beyond the
weak-coupling regime covered in this paper. It
would be particularly interesting to see which
aspects of our solution persist away from weak
coupling, and what are the new qualitative
features that are introduced as the electron-phonon
coupling is increased. The study of stronger
couplings along these lines is left for future
work.

\begin{acknowledgments}
This work was supported in part by the US-Israel
Binational Science Foundation through grant no.
2008440. YV is grateful to the condensed matter
theory group at Rutgers university for their kind
hospitality during the early stages of this work.
\end{acknowledgments}

\appendix

\section{Solution of the scattering-state operators}
\label{app_scat_state}

In this Appendix, we detail the solution of the
scattering-state operators for the different cases
covered in the main text. Altogether three cases
are considered:
(i) a level at resonance with the Fermi energy, i.e.,
    $\epsilon_d = 0$,
(ii) a level off-resonance with the Fermi level, i.e.,
     $\epsilon_d \neq 0$, and
(iii) a local phonon with the shifted frequency
      $\omega_1 = \omega_0 + \delta \omega$.
As emphasized in the main text, in the latter case
we are interested in expanding the scattering-state
operators corresponding to the frequency $\omega_1$
in terms of those corresponding to the original
frequency $\omega_0$.

\subsection{A level at resonance with the Fermi energy}
\label{App-e_d=0}

We begin with a level at resonance with the Fermi
energy, corresponding to the Hamiltonian of
Eq.~(\ref{H_bosonic}) with $\tilde{\epsilon}_d = 0$.
Our objective is to solve the Lippmann-Schwinger
equation
\begin{equation}
\bigl [
        \alpha^{\dagger}_k , {\cal H}
\bigr ] = - \epsilon_k \alpha^{\dagger}_k
          + i \eta \bigl (
                           a^{\dagger}_k
                           - \alpha^{\dagger}_k
                   \bigr ) ,
\label{LSE}
\end{equation}
where $\eta \to 0^{+}$ is a positive infinitesimal.
To this end, we employ the methodology developed in
Ref.~\onlinecite{SH98}. Introducing the Liouville
operator ${\cal L} \hat{O} =
\bigl [ \hat{O}, {\cal H} \bigr ]$, Eq.~(\ref{LSE})
is rewritten in the form
\begin{equation}
\bigl (
        {\cal L} + \epsilon_k + i \eta
\bigr ) \alpha^{\dagger}_k
= i \eta a^{\dagger}_k ,
\end{equation}
which has the formal solution
\begin{equation}
\alpha^{\dagger}_k =
       \frac{i \eta}
            {{\cal L} + \epsilon_k + i \eta}
       a^{\dagger}_k .
\label{LSE-formal-solution}
\end{equation}
Next we divide the Hamiltonian ${\cal H}$ into
three parts,
\begin{eqnarray}
{\cal H}_0 &=&
      \sum_{k > 0}
            \epsilon_k a^{\dagger}_k a_k ,
\label{divide-H0} \\
{\cal H}_1 &=& \omega_0 b^{\dagger} b ,
\\
{\cal H}_2 &=& \lambda ( b^{\dagger} + b )
      \sum_{q > 0} \xi_q
                   \left (
                           a_q + a^{\dagger}_q
                   \right) ,
\label{divide-H2}
\end{eqnarray}
and associate each Hamiltonian term with its own
Liouville operator: ${\cal L}_n \hat{O} =
\bigl [ \hat{O}, {\cal H}_n \bigr ]$ ($n = 0, 1, 2$).
Using the operator identity
\begin{equation}
\frac{1}{{\cal L} + \epsilon_k + i\eta} =
     \left [
             1 - \frac{1}{{\cal L} + \epsilon_k + i\eta}
                 \left (
                         {\cal L}_1 + {\cal L}_2
                 \right )
     \right ]
     \frac{1}{{\cal L}_0 + \epsilon_k + i\eta}
\label{Operator_identity}
\end{equation}
in combination with
\begin{equation}
\left (
        {\cal L}_0 + \epsilon_k + i\eta
\right ) a^{\dagger}_k = i \eta a^{\dagger}_k ,
\end{equation}
\begin{equation}
{\cal L}_1 a^{\dagger}_k = 0 ,
\end{equation}
and
\begin{equation}
{\cal L}_2 a^{\dagger}_k =
     -\lambda \xi_k ( b^{\dagger} + b ) ,
\end{equation}
Eq.~(\ref{LSE-formal-solution}) is recast in the form
\begin{equation}
\alpha^{\dagger}_k = a^{\dagger}_k
       + \lambda \xi_k
         \frac{1}{ {\cal L} + \epsilon_k + i \eta}
         \bigl (
                 b^{\dagger} + b
         \bigr ) .
\label{LSE-1}
\end{equation}

Equation~(\ref{LSE-1}) features two unknown quantities,
\begin{equation}
A_k =  \frac{1}{ {\cal L} + \epsilon_k + i \eta}
       b^{\dagger}
\;\;\;\; {\rm and} \;\;\;\;
B_k = \frac{1}{ {\cal L} + \epsilon_k + i \eta} b .
\label{A_k-B_k-def}
\end{equation}
Our next goal is to explicitly compute these two
operators by expressing them as the solution of
two coupled linear equations. Once at hand, the
scattering-state operator is simply given by
$\alpha^{\dagger}_k = a^{\dagger}_k +
\lambda \xi_k (A_k + B_k)$.

To find $A_k$ and $B_k$ we resort once again to the
operator identity of Eq.~(\ref{Operator_identity}).
Carrying out the relevant commutators one obtains
the pair of equations
\begin{equation}
(\epsilon_k - \omega_0 + i\eta) A_k =
      b^{\dagger}
      + \lambda \sum_{q > 0} \xi_q
        \frac{1}{{\cal L} + \epsilon_k + i\eta}
             ( a_q^{\dagger} + a_q ) ,
\label{A_k-1}
\end{equation}
\begin{equation}
(\epsilon_k + \omega_0 + i\eta) B_k = b
      - \lambda \sum_{q > 0} \xi_q
        \frac{1}{{\cal L} + \epsilon_k + i\eta}
             ( a_q^{\dagger} + a_q ) .
\label{B_k-1}
\end{equation}
Applying yet again the operator identity of
Eq.~(\ref{Operator_identity}) to the right-most
term in Eqs.~(\ref{A_k-1}) and (\ref{B_k-1})
one arrives at
\begin{equation}
(\epsilon_k - \omega_0 + i\eta) A_k =
      b^{\dagger}
      + \Sigma(\epsilon_k + i\eta) (A_k + B_k)
      + C_k ,
\label{A_k-2}
\end{equation}
\begin{equation}
(\epsilon_k + \omega_0 + i\eta) B_k = b
      - \Sigma(\epsilon_k + i\eta) (A_k + B_k)
      - C_k ,
\label{B_k-2}
\end{equation}
where
\begin{equation}
\Sigma(z) = \lambda^2
      \sum_{q > 0} \xi_q^2
           \left (
                   \frac{1}{z - \epsilon_q}
                   - \frac{1}{z + \epsilon_q}
           \right )
\end{equation}
is the phononic self-energy and $C_k$ equals
\begin{equation}
C_k = \lambda
      \sum_{q > 0} \xi_q
           \left (
                   \frac{a^{\dagger}_q}
                        {\epsilon_k - \epsilon_q + i\eta}
                   - \frac{a_q}
                          {\epsilon_k + \epsilon_q + i\eta}
           \right ) .
\end{equation}
Here in deriving Eqs.~(\ref{A_k-2}) and (\ref{B_k-2})
we made use of the fact that
\begin{equation}
\frac{1}{{\cal L}_0 + \epsilon_k + i\eta}
        \left (
                \begin{array}{c}
                      a^{\dagger}_q \\
                      a_q
                \end{array}
        \right ) =
        \frac{1}{\epsilon_k \mp \epsilon_q + i\eta}
        \left (
                \begin{array}{c}
                      a^{\dagger}_q \\
                      a_q
                \end{array}
        \right) .
\end{equation}
Finally, introducing the $2\!\times\!2$ phononic
Green function
\begin{equation}
\hat{G}(z) =
     \left [
             \begin{array}{cc}
                   z - \omega_0 - \Sigma(z) & -\Sigma(z)
                   \\ \\
                   - \Sigma(z) & -z - \omega_0 - \Sigma(z)
             \end{array}
     \right ]^{-1} ,
\label{Phonon-GF}
\end{equation}
Eqs.~(\ref{A_k-2}) and (\ref{B_k-2}) are rewritten in
the compact form
\begin{equation}
\sigma_z \hat{G}^{-1}(\epsilon_k + i\eta)
    \left (
            \begin{array}{c}
               A_k \\
               B_k
            \end{array}
    \right )
    = \left (
              \begin{array}{c}
                 b^{\dagger} + C_k \\
                 b - C_k
              \end{array}
      \right ) ,
\end{equation}
whose solution is
\begin{equation}
    \left (
            \begin{array}{c}
               A_k \\
               B_k
            \end{array}
    \right )
    = \hat{G}(\epsilon_k + i\eta) \sigma_z
      \left (
              \begin{array}{c}
                 b^{\dagger} + C_k \\
                 b - C_k
              \end{array}
      \right ) .
\label{A_k-B_k-solution}
\end{equation}
Here $\sigma_z$ is the Pauli matrix. The
scattering-state operators specified in
Eq.~(\ref{eq_sc_state}) are obtained by
combining Eqs.~(\ref{LSE-1}), (\ref{A_k-B_k-def}),
and (\ref{A_k-B_k-solution}). Note that the function
$g(z)$ defined in Eq.~(\ref{eq_g_defined}) is
simply minus the determinant of $\hat{G}^{-1}(z)$.

\subsection{Extension to nonzero $\epsilon_d$}
\label{App-nonzero-e_d}

The case of a level off-resonance with the Fermi
energy can, in principle, be treated using the
same machinery as the one employed for
$\epsilon_d = 0$. We, however, shall present
a more concise derivation that makes use of
the scattering-state operators obtained for
$\epsilon_d = 0$. As in the main text, the
notation $\alpha^{\dagger}_k$ will be reserved
for the scattering-state operators when
$\epsilon_d = 0$ while the new operators for
$\epsilon_d \neq 0$ are denoted by
$\beta^{\dagger}_k$.

Our starting point is the formal solution
\begin{equation}
\beta^{\dagger}_k =
       \frac{i \eta}
            {{\cal L} + \epsilon_k + i \eta}
       a^{\dagger}_k ,
\end{equation}
where ${\cal L}$ pertains this time to the full
Hamiltonian of Eq.~(\ref{H_bosonic}) with
$\tilde{\epsilon}_d \neq 0$. Using the notations
of Eqs.~(\ref{divide-H0})--(\ref{divide-H2}),
we divide the full Hamiltonian into two parts:
${\cal H}_{\epsilon_d = 0} = {\cal H}_0 +
{\cal H}_1 + {\cal H}_2$ and
\begin{equation}
{\cal H}_{\epsilon_d} = \tilde{\epsilon}_d
      \sum_{q>0} \xi_q
      \left(
             a_q + a^{\dagger}_q
      \right) .
\label{beta-formal-solution}
\end{equation}
Denoting the corresponding Liouville operators
by ${\cal L}_{\epsilon_d = 0}$ and
${\cal L}_{\epsilon_d}$, respectively, we
employ the operator identity
\begin{equation}
\frac{1}{{\cal L} + \epsilon_k + i\eta} =
     \left [
              1 -
              \frac{1}{{\cal L} + \epsilon_k + i\eta}
              {\cal L}_{\epsilon_d}
     \right ]
     \frac{1}
     {{\cal L}_{\epsilon_d = 0} + \epsilon_k + i\eta}
\end{equation}
to rewrite Eq.~(\ref{beta-formal-solution}) in the
form
\begin{equation}
\beta^{\dagger}_k =
     \left [
              1 -
              \frac{1}{{\cal L} + \epsilon_k + i\eta}
              {\cal L}_{\epsilon_d}
     \right ]
     \frac{i \eta}
     {{\cal L}_{\epsilon_d = 0} + \epsilon_k + i\eta}
     a^{\dagger}_k .
\end{equation}
Recognizing that
\begin{equation}
\frac{i \eta}
     {{\cal L}_{\epsilon_d = 0} + \epsilon_k + i\eta}
     a^{\dagger}_k = \alpha^{\dagger}_k ,
\end{equation}
we thus arrive at
\begin{equation}
\beta^{\dagger}_k = \alpha^{\dagger}_k
    - \frac{1}{{\cal L} + \epsilon_k + i\eta}
              {\cal L}_{\epsilon_d}
              \alpha^{\dagger}_k .
\label{beta-expressed-via-alpha}
\end{equation}

Since $\alpha^{\dagger}_k$ and ${\cal H}_{\epsilon_d}$
are both linear in the original bosonic degrees of
freedom, their commutator is a simple $c$-number:
\begin{eqnarray}
{\cal L}_{\epsilon_d} \alpha^{\dagger}_k &=&
    - \tilde{\epsilon}_d \xi_k
    \left [
            1 +
            g(\epsilon_k + i\eta)
            2 \omega_0 \Sigma(\epsilon_k + i\eta)
    \right ]
\nonumber \\
    &=& - \tilde{\epsilon}_d \xi_k
          g(\epsilon_k + i\eta)
          \left (
                  \epsilon_k^2 - \omega_0^2
          \right ) .
\end{eqnarray}
Consequently, using
Eq.~(\ref{beta-expressed-via-alpha}),
\begin{equation}
\beta^{\dagger}_k = \alpha^{\dagger}_k
    + \tilde{\epsilon}_d \xi_k
      \frac{\epsilon_k^2 - \omega_0^2}
           {\epsilon_k + i\eta}
      g(\epsilon_k + i\eta) ,
\end{equation}
which is precisely Eq.~(\ref{beta-via-alpha}).

\subsection{Change in frequency from $\omega_0$
            to $\omega_1 = \omega_0 + \delta \omega$}

Lastly, we wish to expand the scattering-state
operators $\gamma^{\dagger}_k$ corresponding to the
Hamiltonian ${\cal H}' = {\cal H} + \delta {\cal H}$
in terms of those corresponding to ${\cal H}$ alone
(i.e., the $\alpha_k$'s and $\alpha^{\dagger}_k$'s
derived in Sec.~\ref{App-e_d=0}). Here ${\cal H}$
is the full Hamiltonian of Eq.~(\ref{H_bosonic})
with $\tilde{\epsilon}_d = 0$ and $\delta {\cal H}$
equals
\begin{equation}
\delta {\cal H} = \delta \omega b^{\dagger} b .
\end{equation}

As in the previous subsection, we begin from the
formal solution
\begin{equation}
\gamma^{\dagger}_k =
        \frac{i\eta}
             {{\cal L}' + \epsilon_k + i\eta}
        a^{\dagger}_k ,
\label{LSE-formal-solution-g}
\end{equation}
where ${\cal L}'$ is the Liouville operator
associated with ${\cal H}'$. Denoting the
Liouville operators corresponding to ${\cal H}$
and $\delta {\cal H}$ by ${\cal L}$ and
${\cal L}_{\delta \omega}$, respectively, we
make use of the operator identity
\begin{equation}
\frac{1}{{\cal L}' + \epsilon_k + i\eta} =
     \left [
             1
             - \frac{1}{{\cal L}' + \epsilon_k + i\eta}
               {\cal L}_{\delta \omega}
     \right ]
     \frac{1}{{\cal L} + \epsilon_k + i\eta}
\label{op-identity-subsec3}
\end{equation}
to rewrite Eq.~(\ref{LSE-formal-solution-g}) in the
form
\begin{equation}
\gamma^{\dagger}_k =
       \left [
                1 -
                \frac{1}{{\cal L}' + \epsilon_k + i\eta}
                {\cal L}_{\delta \omega}
       \right ]
       \frac{i \eta}
       {{\cal L} + \epsilon_k + i\eta}
       a^{\dagger}_k .
\end{equation}
Recognizing once again that
\begin{equation}
\frac{i \eta}
     {{\cal L} + \epsilon_k + i\eta}
     a^{\dagger}_k = \alpha^{\dagger}_k ,
\end{equation}
we arrive at
\begin{equation}
\gamma^{\dagger}_k = \alpha^{\dagger}_k
      - \frac{1}{{\cal L}' + \epsilon_k + i\eta}
                {\cal L}_{\delta \omega}
                \alpha^{\dagger}_k ,
\label{modified-LSE}
\end{equation}
which is analogous to
Eq.~(\ref{beta-expressed-via-alpha}) of the
previous subsection.

It is straightforward to confirm that
Eq.~(\ref{modified-LSE}) is equivalent to the
modified Lippmann-Schwinger equation
\begin{equation}
[\gamma^{\dagger}_k, {\cal H}']
     = -\epsilon_k \gamma^{\dagger}_k
     + i\eta (\alpha^{\dagger}_k - \gamma^{\dagger}_k) .
\end{equation}
Its usefulness stems from the fact that it allows
one to directly expand $\gamma^{\dagger}_k$ in
terms of the $\alpha_q$'s and $\alpha^{\dagger}_q$'s
without resorting to the separate expansions of
$\gamma^{\dagger}_k$ and $\alpha^{\dagger}_k$ in
terms of the original bosonic degrees of freedom.
Indeed, using Eq.~(\ref{b_via_sc_state}) and its
Hermitian conjugate one has that
\begin{equation}
{\cal L}_{\delta \omega} \alpha^{\dagger}_k =
      \delta \omega
      \lambda \xi_k g(\epsilon_k + i \eta)
      \left [
              (\epsilon_k - \omega_0) b
              - (\epsilon_k + \omega_0) b^{\dagger}
      \right ] ,
\end{equation}
such that
\begin{equation}
\gamma^{\dagger}_k = \alpha^{\dagger}_k
      + \delta \omega
        \lambda \xi_k g(\epsilon_k + i \eta)
        \left [
                 \epsilon_k A'_k
                 + \omega_0 B'_k
        \right ]
\label{gamma-via-A-and-B}
\end{equation}
with
\begin{equation}
A'_k = \frac{1}{{\cal L}' + \epsilon_k + i\eta}
      \left (
              b^{\dagger} - b
      \right )
\label{A_k-subsec3}
\end{equation}
and
\begin{equation}
B'_k = \frac{1}{{\cal L}' + \epsilon_k + i\eta}
      \left (
              b^{\dagger} + b
      \right ) .
\label{B_k-subsec3}
\end{equation}

Similar to the derivation in Sec.~\ref{App-e_d=0},
$A'_k$ and $B'_k$ are computed by expressing them
as the solution of two coupled linear equations,
obtained by applying the operator identity of
Eq.~(\ref{op-identity-subsec3}) to each of
Eqs.~(\ref{A_k-subsec3}) and (\ref{B_k-subsec3}).
After some lengthy but straightforward algebra
one obtains
\begin{equation}
\hat{M}(\epsilon_k + i\eta)
    \left (
            \begin{array}{c}
                   A'_k \\ B'_k
            \end{array}
    \right )
=   \left (
            \begin{array}{c}
                   \mu_k \\ \nu_k
            \end{array}
    \right ) ,
\label{coupled-eqns-for--and-B}
\end{equation}
with
\begin{equation}
\hat{M}(z) =
     \left [
             \begin{array}{cc}
                   1 - \frac{\delta \omega}{\omega_0}
                       \left (
                                z^2 g(z) - 1
                       \right ) &
                   - \delta\omega z g(z)
                   \\ \\
                   - \delta\omega z g(z) &
                   1 - \delta\omega \omega_0 g(z)
             \end{array}
     \right ] ,
\end{equation}
\begin{equation}
\mu_k = 2 \lambda
    \sum_{q > 0} \xi_q \epsilon_q
         \left [
                 \frac{g(\epsilon_q - i\eta)}
                      {\epsilon_k - \epsilon_q + i\eta}
                 \alpha^{\dagger}_q
               - \frac{g(\epsilon_q + i\eta)}
                      {\epsilon_k + \epsilon_q + i\eta}
                 \alpha_q
         \right ] ,
\end{equation}
and
\begin{equation}
\nu_k = 2 \omega_0 \lambda
    \sum_{q > 0} \xi_q
         \left [
                 \frac{g(\epsilon_q - i\eta)}
                      {\epsilon_k - \epsilon_q + i\eta}
                 \alpha^{\dagger}_q
               + \frac{g(\epsilon_q + i\eta)}
                      {\epsilon_k + \epsilon_q + i\eta}
                 \alpha_q
         \right ] .
\end{equation}
Inverting the matrix $\hat{M}(\epsilon_k + i\eta)$
to extract $A'_k$ and $B'_k$ and substituting
the resulting expressions into
Eq.~(\ref{gamma-via-A-and-B}), one recovers
Eq.~(\ref{gamma_k}) for $\gamma^{\dagger}_k$.


\begin{thebibliography}{99}
\bibitem{comment-on-k=0}
         We have omitted here the contribution of
         the $k = 0$ mode of $\partial_x\phi(x)$, as
         it has no effect on our problem of interest.
\bibitem{comment-on-equilibrium-occ}
         Indeed, it is straightforward to
         confirm that the right-hand side
         of Eq.~(\ref{eq_occ_infty}) equals
         $\int_{-\infty}^{\infty} {\rm Im}
         \{ G_{11}(\epsilon + i\eta) \}
         d \epsilon/\pi$, where $G_{11}(z)
         = \ll\!\!b, b^{\dagger}\!\!\gg_z$
         is the corresponding component of
         the phonon Green function of
         Eq.~(\ref{Phonon-GF}). The latter
         integral is nothing but the
         zero-temperature equilibrium phononic
         occupancy, averaged with respect to the
         full Hamiltonian of Eq.~(\ref{H_bosonic})
         with $\tilde{\epsilon}_d$ set to zero.
\bibitem{comment-on-fits}
         In comparing the parameters $\Omega$
         and $\tau_0$ extracted from the different fits
         used throughout the paper, the figures quoted
         for the deviations are exclusively due to
         $\tau_0$. Indeed, the frequency $\Omega$
         agrees to within less than 0.01\% between
         all fits used.
\bibitem{comment-on-identity}
         This identity is established straightforwardly
         by a formal diagrammatic evaluation of
         $\langle b \rangle$.
\bibitem{comment-on-eta}
         The inclusion of $\eta$ does not affect the
         end result of the calculation since each of
         the numerators in Eq.~(\ref{zeta_without_eta})
         vanishes along with its denominator.
\bibitem{comment-on-A_b}
         Due to the terms multiplying
         $|g(\Omega + i\eta)|^2$, the amplitude of
         oscillations $A_b$ is in fact peaked at a
         slightly lower frequency than $\omega$.
\end{thebibliography}
\end{document}